\documentclass[11pt,letter]{article}

\usepackage[margin=1.2in]{geometry}
\usepackage[utf8]{inputenc} 
\usepackage[T1]{fontenc}
\usepackage{url}
\usepackage{ifthen}
\usepackage{cite}
\usepackage{todonotes}                  
\usepackage{color,soul}    
\usepackage[cmex10]{amsmath} 

\usepackage{./rmylay}
\usepackage{authblk}
\newcommand{\x}{\textbf{x}}
\newcommand{\y}{\textbf{y}}

\newcommand{\s}{\textbf{s}}
\newcommand{\z}{\textbf{z}}

\newcommand{\iid}{\emph{i.i.d.}\xspace}

\newcommand{\pn}{{\mathbb P}(\naturals)}
\newcommand{\pnn}{{\mathbb P}(\naturals^n)}
\newcommand{\pni}{{\mathbb P}(\naturals^\infty)}
\newcommand{\ryn}{r^{(y^n)}}
\newcommand{\ryp}{r^{(\y(p))}}

\newcommand{\blankfootnote}[1]{%
  \begingroup
  \renewcommand\thefootnote{}\footnote{#1}%
  \addtocounter{footnote}{-1}%
  \endgroup
  }
\begin{document}
\title{Tail redundancy and its characterization of compression of
  memoryless sources}
\author{M. Hosseini}
\author{N. Santhanam}
\affil{Department of Electrical and Computer Engineering\\ University of Hawaii, Manoa}



\maketitle

\begin{abstract}
  We\blankfootnote{The conference version~\cite{HS19:isit} of this
    paper has appeared in ISIT 2019 with proof outlines rather than
    complete proofs as in this version.} formalize the tail redundancy
  of a collection of distributions over a countably infinite alphabet,
  and show that this fundamental quantity characterizes the asymptotic
  per-symbol minimax redundancy of universally compressing sequences
  generated \iid from a collection $\cP$ of distributions over a
  countably infinite alphabet.

  Contrary to the worst case formulations of universal compression,
  finite single letter minimax (average case) redundancy of $\cP$ does
  not automatically imply that the expected minimax redundancy of
  describing length-$n$ strings sampled \iid from $\cP$ grows
  sublinearly with $n$. Instead, we prove that universal compression
  of length-$n$ \iid sequences from $\cP$ is characterized by how well
  the tails of distributions in $\cP$ can be universally described,
  showing that the asymptotic per-symbol redundancy of \iid strings is
  equal to the tail redundancy.
\end{abstract}
\vspace{1mm}
\section{Introduction}
Universal compression~\cite{dav73} captures the observation that it is
often unreasonable to posit knowledge of the underlying probability
law $p$ generating data.  Rather, one formalizes a setup where the
generating probability law $p$ is unknown, instead compressing data
with just the knowledge that $p$ belongs to a known collection $\cP$
of probability laws, \eg \iid or Markov distributions. Since the
generating law is unknown, we use a single universal probability law
$q$ for the collection $\cP$ that, hopefully, simultaneously encodes
as well as the underlying unknown $p$ as closely as possible.

In the process, the idea is that the universal $q$ and how well it
encodes the data against the true probability laws, captured by
metrics such as \emph{minimax redundancy} should provide insights on
how much information about the generating model we can glean from the
data. We show that the asymptotic minimax redundancy of universally
compressing memoryless sequences from a large, potentially countably
infinite alphabet is captured by the complexity in the tails of the
distributions, which this paper formalizes as the \emph{tail
  redundancy} of a collection of distributions.

A countably infinite alphabet setup coupled with a finite number of
observations often leads to the question of describing novelty,
something that the observations have not yet revealed. The tail
redundancy can be seen as another handle on this puzzle from an
average case universal compression perspective, complementing the work
done on the Good Turing estimators~\cite{Goo53}, which captures the worst case
universal compression formulations~\cite{OSZ03:s+f}.

\paragraph{Universal compression} Universal compression schemes are
applied in several commercial compression algorithms, but as implied
above, their theoretical underpinnings have implications beyond
compression. Metrics used to quantify universal compression
algorithms, in particular, \emph{redundancy}, have interpretations
that lend themselves to regularization~\cite{mdlbook,risbook},
quantifying the information content in observations about the
generating source via the redundancy-capacity
theorem~\cite{CB90,IH72,Gal79,E80,DL80}, while their Bayesian formulations
lead naturally to non-informative priors~\cite{CB94,XB97,xb00}. In
addition, these metrics have been shown to capture the average
reduction of the log compounded wealth in finance and gambling
theory~\cite{BC88}.

Different applications require different formalizations of redundancy,
but this paper focuses on the \emph{average-case} formulation.  For a
probability distribution $q$, the \emph{redundancy} incurred by a collection
$\cP$ of sources over an alphabet $\cX$ is the supremum over all sources $p\in\cP$, of the expected excess codelength of the universal scheme:
\begin{equation}
  \label{eq:avg}
  \sup_{p\in\cP} \EE \bigl[ \log \frac1{q(X)} - \log \frac1{p(X)} \bigr]
  =
\sup_{p\in\cP}\sum_{x\in\cX}p(x)\log\frac{p(x)}{q(x)},
\end{equation}
where the $\cX$-valued random variable $X$ above is distributed
according to $p$.  Alternate worst case formulations measure the
excess codelength of a universal probability law $q$ for the most
inconvenient choice of source and data $x\in\cX$.

We focus now on the optimal value~\eqref{eq:avg} can attain---the
\emph{minimax redundancy} which is the infimum of~\eqref{eq:avg} over
all possible distributions $q$ over $\cX$.  The minimax redundancy has
an elegant statistical interpretation: the minimax redundancy is the
capacity of the channel from $\cP$ to $\cX$---therefore, the amount of
information the observation can provide about the model.  This
equivalence has been well known since~\cite{DL80, Gal79}
when $\cX$ is finite, and was extended to arbitrary alphabets
in~\cite{HO97}.

Leaving formal definitions to Section~\ref{SecDB}, the
redundancy in~\eqref{eq:avg} when $\cX$ is the set of all length-$n$
sequences is termed as the length-$n$ redundancy incurred by a
distribution $q$. We see this distribution $q$ on length-$n$ strings
as the marginal induced on length-$n$ strings by a probability measure
on infinite sequences. For such a probability measure (which we also
denote by $q$ below) on infinite sequences, the asymptotic per-symbol
redundancy,
\[
  \limsup_{n\to\infty} \sup_{p\in\cP}\frac1n\sum_{\x\in\naturals^n}p(\x)\log\frac{p(\x)}{q(\x)},
\]
captures the growth of the (normalized) length-$n$ redundancy.  This
paper specifically addresses this question, characterizing the
asymptotics of the redundancy of compressing sequences of symbols
from a countable alphabet, generated \iid. 

\paragraph{Countably infinite alphabets}
The asymptotics when we have countably infinite supports help provide
insights about scenarios when we have alphabets that are comparable or
even exceed the sample length from which we are learning. This
situation is not uncommon in language modeling, for example, and
different scalings of the sample length and alphabet sizes have been
considered in~\cite{os04:soi,wvk11,SW12} to shed light on different
nuances in this setup. One fundamental aspect of many of these
problems, as mentioned before, is the aspect of describing novelty,
\ie something the finite observation may not have revealed about the
source.

The tail redundancy is another hook to think about describing novelty,
and indeed, it captures the rate at which the minimax redundancy
scales asymptotically. That it should be so is satisfying intuitively,
since it means that we conceptually divide up the description of
sequences into two parts: a part that involves describing novel
symbols that have never been seen prior (which is asymptotically the
dominant contributor to the minimax redundancy) and the description of
statistics of those symbols that we have already seen (which can be
done efficiently). This is indeed reflected in the proofs as well.

A different issue arises from the fact that the asymptotic minimax
redundancy of compression is not as well characterized in the
countably infinite case.  Starting from \cite{dav73}, the case where
$\cX$ is a set of length-$n$ sequences from a set of size $k$, and
$\cP$ is a collection of \iid or Markov probability laws have been
studied extensively. A cursory set of these papers include
\cite{dav81,xb00,ds04} for compression of \iid sequences of sequences
drawn from $k-$sized alphabets, \cite{STW95,Wil98} for context tree
sources, as well as extensive work involving renewal processes
\cite{CsSh96,FS02}, finite state sources \cite{WRF95}, etc.  For the
infinite or large alphabet cases, the results that exist are not as
comprehensive, though this has been approached from multiple
angles. We describe these in the context of this paper in detail in
Section~\ref{s:background}.  \ignore{Given the relevance of the metric
  to a host of insights in regularization and learning in addition to
  compression, it is therefore important to understand the asymptotics
  of redundancy more generally---and} This paper provides the leading
term for the minimax redundancy in general for memoryless sequences.




The specific case, where the asymptotic per-symbol redundancy is zero
is interesting as well. It is well understood that in this scenario,
one can learn from the samples the (unknown) underlying probabilities
of sequences. Asymptotic zero per-symbol redundancy is also sufficient
to be able to learn the marginal distributions as
well, though it is not necessary---see~\cite[Theorem 9]{SAS22:jmlr} for a full
characterization of the learnability of the marginals in the \iid
case.

\ignore{Given its fundamental importance, there has been efforts to
characterize when the asymptotic per-symbol average redundancy is 0,
and more generally, get a better understanding of what the per-symbol
redundancy may be.  Partial results exist, as will be outlined in
detail in Section~\ref{s:background}.}

\paragraph{Summary} In this paper, we obtain a complete single letter characterization for
the asymptotic per-symbol redundancy of length-$n$ strings generated
by \iid sampling, and in the process also settle the open problem of
when a class of \iid sources can be compressed with asymptotically
zero per-symbol redundancy. Section~\ref{SecDB} defines the notion of
redundancies formally, the connection between redundancy and
tightness, and considers results connecting single letter and
asymptotic per-symbol redundancy that will motivate the development of
the paper. Section~\ref{SecRes} introduces the notion of \emph{tail
  redundancy} that is central to the paper, and captures how much
complexity lurks in tails of distributions:
\begin{equation}
  \label{eq:dfn}
\cT(\cP) \ed \limsup_{m\to\infty} \inf_{q} \sup_{p\in\cP} \sum_{x\ge m} p(x) \log\frac{p(x)}{q(x)}.
\end{equation}
Section~\ref{s:rfn} then sets about simplifying the above definition, showing
in several steps that there is a single distribution $q$ satisfying
\[
  \lim_{m\to\infty} \sup_{p\in\cP} \sum_{x\ge m} p(x) \log\frac{p(x)}{q(x)} = \cT(\cP).
\]
We do so by first showing that the limsup in~\eqref{eq:dfn} can be
replaced by a limit, that the limsup and inf can be interchanged (so
it is unnecessary to consider a separate distribution for each $m$),
and that the inf can be replaced by a minimization (there is a
distribution that achieves the tail redundancy). Section~\ref{s:prop}
develops some properties of the tail redundancy: (i) non-negativity,
and (ii) the tail redundancy of finite unions equals the max of tail
redundancies of the components of the union (unlike the single-letter
redundancy of a union of distributions, but like the asymptotic
per-symbol redundancy of unions of classes).

Section~\ref{s:main} uses the material developed to show our main result, that the asymptotic
per-symbol redundancy of \iid classes equals the tail redundancy of
their marginals. That is, for all sets $\cP$ of distributions over $\naturals$,
\[
\limsup_{n\to\infty}\inf_q \sup_{p\in\cP^n} \frac1n{\mathbb E}\log \frac{p(X^n)}{q(X^n)} = \cT(\cP),
\]
where $\cP^n$ is the collection of distributions on $\naturals^n$
obtained as \iid assignments from (\ie products of) the distributions $p\in\cP$. Put
another way, using the redundancy capacity theorem~\cite{Gal79,DL80, MF95, HO97}, the tail redundancy tells us the rate at which we will keep
learning new information about the underlying source from the data.




\section{Prior work} 
\label{s:background}
Consider a collection of all probability measures over a Borel
sigma-algebra over infinite sequences of natural numbers obtained by
\iid sampling from a distribution in $\cP$ over $\naturals$. A
fundamental and natural question that has remained open on this topic
is characterizing the per-symbol asymptotic minimax redundancy of
compressing \iid sequences. It is not even known what would be
necessary and sufficient conditions on $\cP$ so that \iid sequences
from sources in $\cP$ could be compressed with asymptotically zero
per-symbol minimax redundancy.

On the other hand, the question of when sources can be compressed with
asymptotically zero worst case per-symbol redundancy is
settled in work by Boucheron, Garivier and Gassett~\cite{BGG08}. In
the worst case, if a class $\cP$ of distributions over $\naturals$ has
finite worst case redundancy, then length $n$ \iid sequences from
$\cP$ incur asymptotically zero per-symbol redundancy. Of course,
since the average case redundancy is always upper bounded by the worst
case redundancy, we infer that finiteness of the worst case redundancy
of single letter marginals is sufficient to guarantee asymptotically
zero per-symbol average redundancy. On the other hand, finiteness of
single worst case redundancy is not necessary, it is possible to
construct an \iid class whose worst case redundancy of single letter
marginals is infinite, yet the asymptotic per-symbol average
redundancy is 0.

It is also helpful to compare with the single letter characterization
of weak universality. A class of stationary ergodic sources $\cP$ over
sequences of natural numbers is weakly compressible if there is a
distribution $q$ on naturals that incurs only a finite excess
codelength over any single letter marginal of $\cP$.  Note that the
excess codelength of $q$ is finite for each source, but need not be
universally bounded over all sources. It is therefore tempting to
consider whether finiteness of single letter average redundancy
guarantees asymptotic zero per-symbol redundancy. Unfortunately, this
need not be true either (Corollary~\ref{cor:int} in
Section~\ref{s:wc}). While finite single letter average redundancy is
necessary for zero asymptotic per-symbol redundancy, it is not
sufficient.

From another direction, a series of results on grammar based
codes~\cite{KY00}, and in particular~\cite{HY03}, have shown how to obtain
encoding schemes for general stationary ergodic model classes that
incur zero redundancy for all sources. Here the convergence is not
necessarily uniform over the class as in the formulations we have
considered, namely, there is no sup over $p$ in~\eqref{eq:avg}---rather
convergence is considered pointwise for each $p\in\cP$.
It is curious therefore, to see if results on grammar based codes may be
extended to shed light on our problem. But as explained in the Appendix,
approaches from~\cite{HY03} do not lend themselves to a sufficiency
condition even for \iid classes.

In yet another direction, partial results were obtained by Haussler
and Opper~\cite{HO97}, who proved that if a class $\cP$ is not totally
bounded in the Hellinger metric, then the class cannot incur
asymptotically zero per-symbol redundancy. But as we will see,
(Proposition~\ref{prop:hoi} in Section~\ref{SecME}), this result is also incomplete.

\ignore{  Our focus will remain on the
average-case formulations for encoding length-$n$ strings.
This is the first single letter characterization that is both
  necessary and sufficient for strong compressibility in the average
  case to the extent we are aware of.}


\ignore{\paragraph{Bayes Redundancy}
A well known lower bound on average case redundancy relates it to Bayes redundancy of any given prior. This result can be obtained from general minimax theorem. Here we provide the version from~\cite{HO97}.
\bLemma
\label{BayesLem}
Let $\cM(x^n)$ show the set of all probability measure on $X^n$. Denotes elements of $\cP^n$ as $p_{\theta},\theta\in\Theta$. Then Redundancy is lower bounded by Bayes redundancy of any given prior $\pi$ on $\Theta$. i.e. 
\[
R(\cP^n)\geq \inf_{q\in\cM(x^n)} E_{\pi}D(p_\theta||q)
\]
\eLemma
\bLemma[Minimax Theorem~\cite{haussler1997general}]
\label{lMinimax}
Let $\cM(x^n)$ show the set of all probability measure on $X^n$. Denotes elements of $\cP^n$ as $p_{\theta\in\Theta}$.  Then Redundancy is lower bounded by Bayes redundancy of any given prior.
\[
R(\cP^n)=\sup_{\pi\in\cM(\Theta)} \inf_{q\in\cM(x^n)} E_{\pi}D(p_\theta||q)=\sup_{p_\theta\in\cP^n} \inf_{q\in\cM(x^n)}D(p_\theta||q)
\]
\eLemma}

\ignore{\paragraph{Adaptive Coding}
Consider different collections of distribution each of them is universally compressible. Can we find a single encoder for all collections? This is the question of adaptive coding which was answered in~\cite{bontemps2014adaptive}~\cite{boucheron2015adaptive}. 
The idea of adaptive coding is very close to hierarchical universal compression~\cite{mf96}. Since it is not possible to compress a general class of distributions over countable alphabet. In \cite{bontemps2014adaptive}, authors consider an envelope class and proposes an auto censoring code and show that it is adaptive for an envelope class.
In~\cite{boucheron2015adaptive} authors investigate a heavy tail envelope class and  obtain a lower bound on the redundancy of this class.
\paragraph{}
Rest of the paper is organized as follows.  Section~\ref{SecDB} lays out formal
notations and definitions for strong and weak compression and
redundancy along with the summary of previous results. Section~\ref{SecME} connects result of the paper to metric entropy and provides a few examples on that. Section~\ref{SecRes}
states the condition that is both necessary and sufficient for the
strong compression guarantee to hold in Theorem~\ref{Thm1}.}

\section{Definitions and Background}
\label{SecDB}

The following development of universal compression is essentially
standard. However, for formal simplicity in definitions we define
redundancy and its asymptotics by means of measures over infinite
sequences rather than sequences of distributions over various lengths.

Let $\naturals={1,2,\ldots}$ be the set of naturals, $\naturals^*$ be
the collection of all finite strings of naturals, and let $\pn$
($\pnn$ respectively) be the set of all probability distributions over
$\naturals$ ($\naturals^n$ respectively). Let $\cP\subset \pn$ be a
collection of distributions over $\naturals$ and $\cP^n$ be the set of
distributions over length-$n$ sequences, $\naturals^n$, obtained via
\iid assignments from marginals $p\in\cP$ (\ie products
distributions). For all $p$, \iid assignments of probabilities for
finite length strings can be naturally extended to a probability
measure on the Borel sigma-algebra on the natural product topology in
$\naturals^\infty$, see~\eg~\cite[Chapter 2]{JR}\footnote{Since
  $\naturals^\infty$ is not countable, we need to define an
  appropriate sigma-algebra. The one we choose allows us to focus on
  probability assignments on finite strings in $\naturals^*$ alone. We
  note the sigma algebra here for completeness, and will not need
  to delve into further details.}. Let $\cP^\infty$ be the collection
of all such probability measures over infinite length sequences of
$\naturals$ obtained through the above construction. We use the same
symbol $p$ to indicate the probability measure in $\cP^\infty$, or its
marginals---the distributions in $\cP$ or $\cP^n$.

We will use $\pni$ for the set of all probability measures over the
Borel sigma-algebra on the product topology in $\naturals^\infty$ (all
\emph{sequential estimators} as sometimes used in compression
literature), and these probability measures can be specified uniquely
by simply specifying the probabilities they assign on every finite
string of natural numbers.  The term sequential estimator in
compression literature recalls the fact that the induced distributions
on $\pnn$ are consistent, and one can assign probabilities on any
finite string via a sequence of conditional probabilities. Let $q$ be
an arbitrary (not necessarily \iid) probability measure in $\pni$. For
all $n$, the redundancy of $q$ against any $p\in\cP^\infty$ is
\footnote{All the logarithms are in base 2, unless otherwise
  specified.}
\begin{equation}
\label{Rn}
R_n(p,q)=\sum_{x^n\in \naturals^n}p(x^n)\log\frac{p(x^n)}{q(x^n)} \ed D_n(p||q),
\end{equation}
where $D_n()$ above denotes the KL divergence between the distributions on
$\naturals^n$ induced by the measures $p$ and $q$ respectively. 
A collection $\cP^\infty$ is \textit{weakly compressible} if there exists a probability measure $q\in\pni$ such that for all $p\in\cP^\infty$
\[
\lim_{n\to\infty}\frac1{n}R_n(p,q)=0.
\]
A collection $\cP^\infty$ is \textit{strongly compressible} if there
exists a probability measure $q\in\pni$ such that
\[
\lim_{n\to\infty}\sup_{p\in\cP^n}\frac1{n}R_n(p,q)=0.
\]
Define the length-$n$ \emph{per-symbol redundancy} of any $q\in\pnn$ (or completely equivalently, $q\in\pni$) against $\cP^\infty$ 
to be
\[
  R_n(\cP^\infty,q) \ed \sup_{p\in\cP^n}\frac1{n}R_n(p,q)
\]
and the length-$n$ \emph{per-symbol redundancy} of $\cP^\infty$ to be
\begin{equation}
  \label{eq:rn}
  R_n(\cP^\infty) \ed \inf_{q} R_n(\cP^\infty,q) = \inf_q \sup_{p\in\cP^n}\frac1{n}R_n(p,q)
\end{equation}
where the infimum is taken over all $q \in \pni$, or completely
equivalently, $q\in\pnn$ (as is commonly done).  For $n=1$, the
length-1 redundancy, $R_1(\cP^\infty)$ will be called the \emph{single-letter
  redundancy} of $\cP$, and to emphasize the point, we will use $R_1(\cP)$
to denote the case $n=1$. Finally, the \emph{asymptotic, per-symbol
  redundancy of $q\in\pni$ against $\cP^\infty$}, $R(\cP^\infty,q)$,
as well as the \emph{asymptotic, per-symbol redundancy of $\cP^\infty$}, $R(\cP^\infty)$, (see discussion below and in
Appendix~\ref{app:rdetails}) to be
\begin{equation}
  \label{eq:r}
  R(\cP^\infty,q)\ed \limsup_{n\to\infty}\sup_{p\in\cP^n}\frac1{n}R_n(p,q)
  \text{ and } R(\cP^\infty) = \inf_q R(\cP^\infty,q)
\end{equation}
which we show in Appendix~\ref{app:rdetails} to also satisfy
\begin{equation}
  \label{eq:rho}
  R(\cP^\infty) =\limsup_{n\to\infty}\inf_{q\in\pnn} \sup_{p\in\cP^n}\frac1{n}R_n(p,q) = \limsup_{n\to\infty} R_n(\cP^\infty).
\end{equation}
The definition of asymptotic-per symbol redundancy is a minor
technical departure from some of prior literature, but is completely
equivalent while clarifying the following. In certain prior
expositions, as in the equation~\eqref{eq:rho} above, the infimum
over $q$ does not enforce that the length $n$ distributions for
all $n$ be consistent, \ie, marginals of the same probability
measure. Instead, there is seemingly extra freedom allowed, where
different (potentially inconsistent) probability distributions
$\sets{q_n\in\pnn: n\ge 1}$ could be chosen for different $n$. The seeming
additional ``freedom'' in allowing potentially inconsistent
probability distributions is a red herring, and does not yield any
actual advantage, something automatically clarified by the
definition in~\eqref{eq:r}. See Appendix~\ref{app:rdetails}.

We will need the following elementary results on the redundancy.

\bProposition
\label{prop:one}
For all $\cP$ and all numbers $n\ge1$, $\frac1n R_n(\cP^\infty) \le R_1(\cP)$,
and therefore $R(\cP^\infty) <\infty$.
\Proof
See~\cite[Proposition 38]{} for a proof that for all $n\ge1$, $\frac1n R_n(\cP^\infty) \le R_1(\cP)$. The proposition then follows from Appendix~\ref{app:rdetails} that proves that $R(\cP^\infty)=\limsup_{n\to\infty} \frac1n R_n(\cP^\infty)$.
\eProposition

\bProposition
For all $\cP$, $\cR(\cP^\infty)<\infty$ iff $R_1(\cP)<\infty$.
\Proof See~\cite[Corollary 39]{SAS22:jmlr} for a proof.
\eProposition

\ignore{One can consider weak vs strong compressibility as pointwise
convergence in contrast to uniform convergence. While weak
compressibility needs~\eqref{Rn} to go to zero for each $p$ as
$n\to\infty$, strong compressibility needs the uniform convergence of
$R_n(p,q)$ toward to zero as $n\to\infty$. In fact, we can use Egorov's theorem to connect weak compressibility and strong compressibility using Lemma~\ref{LemEgo}.
%
%
\bLemma[Egorov's Theorem]
\label{LemEgo}
Let $\{f_n(\theta)\}, \theta\in\Theta$ be a sequence of measurable functions on measurable space $(\Theta,\Sigma,\mu)$ where $\mu$ is a finite measure and ${f(\theta)}$ be a measurable functions on this space. If $\{f_n(\theta)\}$ converges to ${f(\theta)}$ pointwise, then for every $\epsilon>0$, there is a subset $B\subset\Theta$ such that $\mu(B)<\epsilon$ and $\{f_n(\theta)\}$ converges to ${f(\theta)}$ uniformly on $B^c=\Theta-B$.\\
\eLemma}

\ignore{\subsection{Strong Compression Over Finite Alphabet}
In \cite{dav73} and \cite{dav81}, Davisson and Davisson \textit{et al.} proved that an arbitrary collection of stationary ergodic distributions over finite alphabet are weakly compressible, but it is strongly compressible if and only if,
\[
\lim_{n\to\infty} \frac{1}{n}\sum_{X^n}p(X^n)\log\frac{1}{p(X^n)}=H(p) \quad [\text{uniformly}]
\]}

\subsection{Tightness}
\newcommand{\prob}{{\mathbb P}}
A collection $\cP\subset \pn$ of distributions on $\naturals$ is defined to be
\emph{tight} if for all $\gamma>0$, there is a number $N_\gamma$ such that
\[
  \sup_{p\in \cP} \prob(X_p> N_\gamma) < \gamma
\] 
where $X_p$ above is a random variable distributed according to $p$. 
\bLemma 
\label{LemTightness}
Let $\cP\subset \pn$ be a class of distributions on $\naturals$ with
finite single letter redundancy, namely $R_1<\infty$. Then $\cP$ is tight.  
\Proof This is a well known folk theorem, see~\cite{HO97,HS19:isit,SAS22:jmlr}
for three separate proofs.
\eLemma

The converse is not necessarily true. Tight collections need not have finite single
letter redundancy as the following example demonstrates.

\paragraph{Construction} Consider the following collection $\cI$ of
distributions over $\naturals$.  First partition the set of naturals
 into the sets $T_i$, $i\in\naturals$, where
\[
T_i = \sets{2^i\upto 2^{i+1}-1}.
\]
Note that $|T_i|=2^i$.
Now, $\cI$ is the collection of all possible distributions that can be formed as
follows---for all $i\in \integersp$, pick exactly one element of $T_i$ and assign 
probability $1/((i+1)(i+2))$ to the element of $T_i$ chosen.
Note that the set $\cI$ is uncountably infinite.\hfill$\Box$

\bCorollary
The set $\cI$ of distributions is tight. 
\Proof For all $p\in\cI$,
\[
\sum_{\substack{x\ge 2^k\\x\in\integersp}} p(x) = \frac1{k+1},
\]
namely, all tails are uniformly bounded over the collection $\cI$. Put another way, for all $\delta>0$ and all distributions
$p\in\cI$,
\[
F_p^{-1}(1-\delta) \le 2^{\frac1\delta}.\eqed
\]
\eCorollaryp

On the other hand,
\bProposition
The collection $\cI$ does not have finite redundancy. 
\Proof
Suppose $q$ is any distribution over $\integersp$. We will show that 
$\exists p \in \cI$ such that 
\[
\sum_{\substack{x\in\integersp}} p(x) \log \frac{p(x)}{q(x)} 
\]
is not finite. Since the entropy of every $p\in\cI$ is finite, we just have to
show that for any distribution $q$ over $\integersp$, there $\exists p \in \cI$
such that
\[
\sum_{\substack{x\in\integersp}} p(x) \log \frac{1}{q(x)}
\]
is not finite.

Consider any distribution $q$ over $\integersp$. Observe that for all
$i$, $|T_i|=2^i$. It follows that for all $i$ there is $x_i\in T_i$ such that
\[
q(x_i) \le \frac1{2^i}.
\]
But by construction, $\cI$ contains a distribution $p^*$ that has for its support
$\sets{x_i:i\in\integersp}$ identified above. Furthermore $p^*$ assigns
\[
p^*(x_i)=\frac1{(i+1)(i+2)} \qquad\forall \,i\in \integersp.
\]
The KL divergence from $p^*$ to $q$ is not finite and the Lemma follows since $q$ is arbitrary.  \eProposition
\subsection{Single letter and asymptotic per-symbol redundancy}
\label{s:wc}
Although arbitrary collections of stationary ergodic distributions
over finite alphabets are weakly compressible, Kieffer \cite{Kie78}
showed the collection of all \iid distributions over $\naturals$
is not even weakly compressible. Indeed, here the finiteness of
single letter redundancy is sufficient for weak compressibility.
Any collection of stationary ergodic measures
over infinite sequences is weakly compressible if $R_1<\infty$.

$R_1$ being finite, however, is not sufficient for strong compression
guarantees to hold even when while dealing with \iid sampling. We
reproduce the following Example \ref{exm1} from~\cite{HS14} to
illustrate the pitfalls with strong compression, and to motivate the
notion of \emph{tail redundancy} that will be central to our main
result.  Proposition~\ref{prop1} shows that the collection in the
Example below has finite single letter redundancy, but Proposition
\ref{prop2} shows that its length $n$ redundancy does not diminish to
zero as $n\to\infty$.

\bExample 
\label{exm1}
Partition the set $\naturals$ into
$T_i=\sets{ 2^i\upto 2^{i+1}-1}$, $i\in\naturals$.  Recall that $T_i$
has $2^i$ elements. For all $n\ge 1$, let $1\le j\le 2^{n}$
and let $p_{n,j}$ be a distribution on $\naturals$ that
assigns probability $1-\frac1n$ to the number 1 (or equivalently, to
the set $T_0$), and $\frac1n$ to the $j$th smallest element of
$T_{n}$, namely the number $2^{n}+j-1$. $\cB$
(mnemonic for binary, since every distribution has at support of size
2) is the collection of distributions $p_{n,j}$ for all
$n>0$ and $1\le j \le 2^{n}$.  $\cB^\infty$ is the set
of measures over infinite sequences of numbers corresponding to \iid
sampling from $\cB$.
\eExample

We first verify that the single letter redundancy of $\cB$ is finite. 
\bProposition
\label{prop1}
Let $q\in\pn$ be a distribution that assigns $q(T_i)=\frac1{(i+1)(i+2)}$ and
for all $j\in T_i$, 
\[
q(j|T_i) = \frac1{|T_i|}.
\]
Then
\[
\sup_{p\in\cB}\sum_{x\in\naturals} p(x) \log \frac{p(x)}{q(x)}\le 2.\eqed
\]
\ePropositionp

However, the redundancy of compressing length-$n$ sequences from $\cB^\infty$ scales
linearly with $n$.

\bProposition
\label{prop2}
 For all $n\in\naturals$,
\[
  \inf_{q} \sup_{p\in B^\infty}  \frac 1n E_p \log \frac{p(X^n)}{q(X^n)} \ge \Paren{1-\frac1e} -\frac1n h\Paren{\frac1e},
\]
where $h(x) = -x\log x -(1-x) \log (1-x)$ is the binary entropy
function and the infimum is over all distributions over $\naturals^n$.
\Proof Let the set $\sets{1^n}$ denote a set containing a length-$n$
sequence of only ones.
For all $n$, define $2^n$ pairwise disjoint sets $S_i$ of $\naturals^n$, $1\le i\le 2^n$, where
\[
S_i = \sets{1, 2^n+i-1}^n -\sets{1^n}
\]
is the set of all length-$n$ strings containing at most two numbers ($1$ and
$2^n+i-1$) and at least one occurrence of $2^n+i-1$. 
Clearly, for distinct $i$
and $j$ between 1 and $2^n$,  $S_i$ and $S_j$ are disjoint.  Furthermore, the
measure $p_{\frac1n,i}\in\cB^\infty$ assigns $S_i$ the probability
\[
p_{\frac1n,i}(S_i) = 1-\Paren{1-\frac1n}^n > 1-\frac1e.
\]
Since there are $2^n$ pairwise disjoint sets $S_i$, no matter what the
universal distribution $q$ over length-$n$ sequences, there must be a
set $S_j$ such that
\[
  q(S_j) \le 2^{-j}.
\]
Therefore, the redundancy incurred by $q$ against $p_{\frac1n,j}$ is 
\[
  E_{p_{\frac1n,j}} \log \frac{p_{\frac1n,j}(X^n)}{q(X^n)} \ge
    p_{\frac1n,j}(S_j) \log \frac{p_{\frac1n,j}(S_j)}{q(S_j)}
    + p_{\frac1n,j}(S_j^c) \log \frac{p_{\frac1n,j}(S_j^c)}{q(S_j^c)}
\]
which is in turn lower bounded by
\[
\Paren{1-\frac1e}\log 2^n - h\Paren{\frac1e}
= n\Paren{1-\frac1e} -h\Paren{\frac1e}.
\]  \eProposition

We will see later that the asymptotic
per-symbol redundancy of $\cB^\infty$ equals 1.

\ignore{For all $\delta>0$, let $T_{p,\delta}$ be the set of all elements in the support set of $p$ with probability smaller than or equal to $\delta$, and let $A_{p,\delta}=\naturals-T_{p,\delta}$ ($A_{p,\delta}$ is the complement of $T_{p,\delta}$ in $\naturals$). Then, for a distribution $q$ over $\naturals$, define
\[
\sup_{p\in\cP}\sum_{x\in T_{p,\delta}}p(x)\log\frac{p(x)}{q(x)}
\]
as \textit{tail redundancy}.
 and define
 \[
 \sup_{p\in\cP}\sum_{x\in A_{p,\delta}}p(x)\log\frac{p(x)}{q(x)}
 \]
 as \textit{tail entropy}.}

\bCorollary\label{cor:int} There exists a collection of distributions $\cP\in\pn$ with finite single letter redundancy, yet $\cP^\infty$, the set of
\iid processes with single letter marginals from $\cP$, has 
asymptotic per-symbol redundancy bounded away from 0.  \Proof $\cB$
from Example~\ref{exm1} is one such class.  \eCorollary

It is instructive to compare what happens when we try to describe a
single digit output from an unknown distribution in $\cB$. An
observation we make is that there is no number $m$ such that some
universal distribution over $\naturals$ describes numbers $\ge m$ as
well as the best distribution in $\cB$, in the sense the redundancy
incurred is always bounded below by 1, no matter how large $m$ is.
This is reflected when compressing strings of length $n$---we incur a
heavy penalty against those distributions that contain an element with
probability $\cO(\frac1n)$.  It is a different set of distributions
that hit us for different sequence lengths, and this does not stop no
matter how large the sequence length becomes. This is the essence of
the problem in Proposition~\ref{prop2}, and what motivates our
definition of tail redundancy in Section~\ref{SecRes}.

\ignore{The collection of monotone distributions with finite entropy is known to be weakly compressible. We now use 
Lemma~\ref{LemTightness} to verify that it is not strongly compressible.
\bExample 
\label{exm2}
Let $\mathcal{M}$ be the collection of monotone distributions over
$\naturals$ with finite entropy.  Let $\mathcal{M}^\infty$ be the set
of all \iid processes obtained from distributions in $\mathcal{M}$.
For all $p\in\mathcal{M}$ and all numbers $n$, we have
\[
p(n)\leq\frac1{n}.
\]
So if $q(n) = \frac6{\pi^2 n^2}$, then for all $p\in\cM$,
\[
  \sum_{n\ge 1} p(n)\log q(n) \le 
  \log \frac6{\pi^2} + 2\sum_{n\geq 1}p(n)\log n
  \leq
  \log \frac6{\pi^2} + 2\sum_{n\geq 1}p(n)\log \frac1{p(n)}\leq\infty,
\]
and from Kieffer's condition, $\mathcal{M}^\infty$ is weakly compressible.
However, it is easy to verify that $\cM$ is not tight. To see this,
consider the collection $\mathcal{U}$ of all uniform distributions
over finite supports of form $\{m, m+1,\dots, M\}$ for all positive
integers $m$ and $M$ with $m\leq M$. Let $\mathcal{U}^\infty$ be the
set of all \iid processes with one dimensional marginal from
$\mathcal{U}$.  Consider distributions of form
$p'=(1-\epsilon)p+\epsilon q$ where
$q\in \mathcal{U} \cap \mathcal{M}$ is a monotone uniform distribution
and $\epsilon>0.$ The $\ell_1$ distance between $p$ and $q$ is
$\leq 2\epsilon$. For all $M >0$ and $\delta \leq \epsilon$, we can
pick $q\in \mathcal{U}$ over a sufficiently large support such that
$F^{-1}_{p'}(1-\delta)>M$, so $\mathcal{M}$ is not tight.
Since $\mathcal{M}$ is not tight, from Lemma~\ref{LemTightness} its
single letter redundancy is not finite. Hence the length-$n$ redundancy
cannot be finite for any $n$ and $\mathcal{M}^\infty$ is not strongly
compressible.  \eExample}
\subsection{ Asymptotic zero per-symbol redundancy and boundedness in Hellinger metric}
\label{SecME}
To connect single letter redundancy to length$-n$ redundancy, the
authors in~\cite{HO97} obtain partial lower and upper bounds on the
asymptotic per-symbol redundancy using the total-boundedness of the probability
set under the Hellinger metric. This work perhaps comes closest to our results,
and we present both the result, as well as why the result is yet incomplete
in this Section.


\bDefinition[Hellinger Distance]
Let $p_1$ and $p_2$ be two distributions in $\pn$. The Hellinger distance $h$ is defined as 
\[
h^2(p_1,p_2)=\frac12\sum_{x\in\naturals} \left(\sqrt{p_1(x)}-\sqrt{p_2(x)}\right)^2.\eqed
\]
\eDefinitionp


\bDefinition[Totally Bounded Set~\cite{HO97}]
Let $(S,\rho)$ be any complete separable metric space. A partition $\Pi$ of set $S$ is a collection of disjoint Borel subsets of $S$ such that their union is $S$. Then diameter of a subset $A\subset S$ is $d(A)=\sup_{x,y\in A} \rho (x,y)$ and diameter of partition $\Pi$ is supremum of diameters of the sets in the partition. 
For $\delta>0$, let $\mathcal{D}_\delta(S,\rho)$ be the cardinality of the smallest finite partition of $S$ of diameter $\le\delta$. We say $S$ is totally bounded if $\mathcal{D}_\delta(S,\rho)<\infty$ for all $\delta>0$.
\eDefinition

\bLemma [~\cite{HO97}] If length$-n$ redundancy is finite it can grow
at most linearly in $n$. If $(\cP,h)$ is not \textit{totally bounded}
in the Hellinger metric and single letter redundancy is finite then
$\liminf_{n\to\infty}\frac1{n}R_n(\cP^n)$ is bounded away from zero and
$\limsup_{n\to\infty}\frac1{n}R_n(\cP^n)<\infty$.  \Proof See~\cite[part
5, Theorem 4]{HO97}.  \eLemma 

The above is not a complete characterization of when the asymptotic
per-symbol redundancy is bounded away from 0, and the converse of the Lemma
above does not hold. For example, recall the
collection $\cB$ from Example 1. We show below that $\cB$ is a counter-example
that proves that the converse of the above lemma cannot hold.

%
%

\bProposition\label{prop:hoi} The collection $(\cB,h)$ is totally
bounded in the Hellinger metric, but
\[
  \liminf_{n\to\infty}\frac1{n}R_n(\cB^n)>0.
  \]
\Proof To see $\cB$ is totally bounded in the Hellinger metric,
consider the following partition for $\delta>0$. Let
$N= \ceil{\frac3{2\delta^2}}$ and we partition $\cB$ into
$\le N2^N+1$ parts, each with diameter $\le \delta$.  All but the last
part contains exactly one distribution each among $p_{n,j}$ where
$n\le N$ and $1\le j\le 2^n$ (therefore, $\le N2^N$ parts.  The last
part of the partition contains all the other distributions of
$\cB$. The diameter of all but the last part is 0 (since they
are sets with only one distribution each). We bound the Hellinger
distance between any two distributions in the last part by noticing
that for numbers $n_1$ and $n_2$ both $>N$,
\[
  \Paren{\sqrt{1-\frac1{n_1}}-\sqrt{1-\frac1{n_2}}}^2 + \frac1{n_1}+\frac1{n_2}
  \le \biggl|\frac1{n_1}-\frac1{n_2}\biggr| +\frac1{n_1}+\frac1{n_2} \le \frac3{N},
\]
which in turn implies that the Hellinger distance between any pair
of distributions in the last part is bounded by $\sqrt{\frac3{2N}}\le \delta$ by our choice of $N$. The redundancy
result follows from Proposition~\ref{prop2}.  \eProposition

In the development below, we give a complete characterization of the
asymptotic per-symbol redundancy.



\ignore{
\subsection{Sub-additivity}
The idea that finite single letter regret implies finite length$-n$ regret is benefited from sub-additivity of it. Here we propose definition of sub-additivity to make the paper self standing and show that if tail redundancy is zero then the average case redundancy is sub-additive. 
\bDefinition[sub-additive]
A sequence ${a_n}$ is sub-additive if for all $m$ and $n$ in $\naturals$ it satisfies
\[
a_{n+m}\leq a_n+a_m.
\]
\eDefinition
\bLemma [subadditivity of average case redundancy]~\cite{boucheron2009coding}
$R(\cP^n)$ and $\hat{R}(\cP^n)$ both are either infinite or sub-additive. 
\eLemma}

%

%
%
%

\ignore{We can use Fekete's Lemma to conclude that if redundancy (regret) is finite then it converges to single letter redundancy (regret).
\bLemma [Fekete's Lemma]
If function $R(\cP^n)$ is sub-additive then  $\frac{R(\cP^n)}{n}\to R(\cP)$ as $n\to\infty$. 
\eLemma}

\section{Tail Redundancy}
 \label{SecRes}
We will develop a series of tools that will help us better understand
how the per-symbol redundancy behaves in a wide range of large alphabet
cases. In particular, for \iid sources, we completely characterize the
asymptotic per-symbol redundancy in terms of single letter marginals. 
Fundamental to our analysis is the understanding of how much complexity
lurks in the tails of distributions. 

To this end, we define what we call the \emph{tail redundancy}. We assert
the basic definition below, but simplify several nuances around it in
Section~\ref{s:rfn}, eventually settling on a operationally workable 
characterization.

\bDefinition
For a collection $\cP$ of distributions, define for all $m\ge1$
\[
\cT_m(\cP) \ed \inf_{q\in\pn} \sup_{p\in\cP} \sum_{x\ge m} p(x) \log\frac{p(x)}{q(x)},
\]
where the infimum is over all distributions $q$ over
$\naturals$.  We define the \emph{tail redundancy} as
\[
\cT(\cP)\ed \limsup_{m\to\infty} \cT_m(\cP).
\]
The above quantity, $\cT_m(\cP)$ can be negative, and is not then
redundancy of any collection of distributions as is conventionally
understood. However, let  $S_m=\{x\in\naturals: x \geq m\}$ and 
\[
\tilde{\cT}_m(\cP)\ed\inf_{q\in\pn} \sup_{p\in\cP} \Paren{\sum _{x\geq m} p(x) \log \frac{p(x)}{q(x)} + p(S_m)\log \frac1{p(S_m)}}
\]
is always non-negative, and can be phrased in terms of a conventional
redundancy. To see this, let $p'$ be the distribution over numbers in $S_m$ 
obtained from $p$ as $p'(x)=p(x)/p(S_m)$, and note that
\[
\tilde{\cT}_m(\cP)
=\inf_{q\in\pn} \sup_{p\in\cP} p(S_m) D_1\left( p'(x)||q(x)\right).\eqed
\]
\eDefinitionp

While we unravel the above definitions in detail in the next section, we first
note that, as with redundancy, only tight classes can possibly have finite tail redundancy. In general if the single letter redundancy is infinite, so is the
tail redundancy.
\bProposition
\label{prop:tight}
For $\cP\subset\pn$, if $R_1(\cP)=\infty$, 
then for all
$m\in\naturals$ and for all distributions $q\in\pn$,
\[
  \sup_{p\in\cP} \sum_{x\ge m} p(x) \log\frac{p(x)}{q(x)} =\infty.
\]
and therefore, $\cT(\cP)=\infty$. In particular if $\cP$ is not tight,
$\cT(\cP)=\infty$.  \Proof Suppose $R_1(\cP)=\infty$, and there exists
$m\in\naturals$ and a distribution $q_m\in\pn$ and some $M<\infty$
such that
\[
\sup_{p\in\cP} \sum_{x\ge m} p(x) \log\frac{p(x)}{q_m(x)} =M.
\]
Consider the distribution $q_1\in\pn$ that assigns probability $1/(m-1)$ for all
numbers from 1 through $m-1$.
Then the distribution $q=(q_1+q_m)/2$ satisfies
\[
\sup_{p\in\cP} \sum_{x\in\naturals} p(x) \log\frac{p(x)}{q(x)} \le M+\log(m-1)+1,
\]
a contradiction that $R_1(\cP)=\infty$. 
Therefore, we can also conclude that
$\cT_m(\cP) =\infty$ for all $m$, and therefore $\cT(\cP)=\infty$.

For the last part, if $\cP$ is not tight, Lemma~\ref{LemTightness} implies
that the single letter redundancy is infinite, and therefore the tail
redundancy is infinite as well.  \eProposition

\subsection{Operational characterization of tail redundancy}
\label{s:rfn}
We refine the above definitions in several ways. First we prove that
the sequence $\cT_m$ always has a limit and
\begin{equation}
\label{dfn:tr}
\cT(\cP)= \lim_{m\to\infty} \cT_m(\cP)
=\lim_{m\to\infty}\inf_{q_m\in\pn} \sup_{p\in\cP} \sum_{x\ge m} p(x) \log\frac{p(x)}{q_m(x)}.
\end{equation}
Next, we show that the limit and inf above can be interchanged, and in
addition, that a minimizer exists---namely there is always a
distribution over $\naturals$ that achieves the tail redundancy. This
will let us operationally characterize the notions in the definitions
above.

\bLemma
\label{lm:tmni}
$\tilde{\cT}_m(\cP)$ is non-increasing in $m$. 
\Proof
Let $q$ be any distribution over $\naturals$ and as before, $S_m=\{x\in\naturals: x \geq m\}$.
We show that 
\[
\sup_{p\in\cP} \Paren{\sum _{x\geq m} p(x) \log \frac{p(x)}{q(x)} + p(S_m)\log \frac1{p(S_m)}}\ge \tilde{\cT}_{m+1}(\cP),
\]
thus proving the lemma.

To proceed, note that without loss of generality we can assume
$\sum_{x\geq m}q_m(x)=1$. For $x\ge m+1$, let
\begin{equation}
  \label{eq:qm}
  q'(x)=\frac{q_m(x)}{\sum_{x\geq m+1}q_m(x)}=\frac{q_m(x)}{1-q_m(m)}.
\end{equation}
We have
\begin{align*}
  \tilde{\cT}_m
  &=\sup_{p\in\cP}
    \biggl(
    \sum_{x\geq m}
    p(x) \log \frac{p(x)}{q(x)}+p(S_m)\log \frac1{p(S_m)}\biggr)\\
  &=\sup_{p\in\cP}
    \biggl(
    p(m)\log \frac{p(m)}{q_m(m)}
    +
    \sum_{x\geq m+1}
    p(x) \log \frac{p(x)}{q_m(x)}
    +
    p(S_m)\log \frac1{p(S_m)}\biggr)\\
  &\aeq{(a)}
    \sup_{p\in\cP}
    \left(
    p(m)\log \frac{p(m)}{q_m(m)}
    +p(S_{m+1})\log\frac1{1-q_m(m)}+
    \sum_{x\geq m+1}p(x)\log \frac{p(x)}{q'(x)}
    \right.\\
  &\qquad\qquad\qquad
    \left.
    +
    p(S_m)\log \frac1{p(S_m)}
    \right)\\
  &\aeq{(b)}
    \sup_{p\in\cP}
    \biggl(
    p(m)\log \frac{p(m)/p(S_m)}{q_m(m)}
    +p(S_{m+1})\log\frac{1/p(S_{m})}{1-q_m(m)}
    +
    \sum_{x\geq m+1}
    p(x)\log \frac{p(x)}{q'(x)}
    \biggr)\\
  &=
    \sup_{p\in\cP}
    \biggl(
    p(m)\log \frac{p(m)/p(S_m)}{q_m(m)}
     +p(S_{m+1})\log\frac{p(S_{m+1})/p(S_{m})}{1-q_m(m)}
     +
     \sum_{x\geq m+1}p(x)\log \frac{p(x)}{q'(x)}\\
   &\qquad\qquad\qquad+p(S_{m+1})\log\frac1{p(S_{m+1})}\biggr)\\
  &\aeq{(c)}                           
    \sup_p
    \biggl[
    p(S_m)
    D_1\biggl(
    B\Paren{
    \frac{p(m)}{p(S_m)}}
    ||
    B(q_m(m))\biggr)
    +\sum_{x\geq m+1}
    p(x)\log\frac{p(x)}{q'(x)}
    +p(S_{m+1})\log\frac1{p(S_{m+1})}
    \biggr]\\
  &\geq \tilde{\cT}_{m+1}(\cP),
 \end{align*}
where in $(a)$, we use~\eqref{eq:qm} for $x\ge m+1$, in $(b)$ we
absorb the last term into the first two terms of the prior equation,
noting that
$p(S_m) -p(m) = p(S_{m+1})$, and in $(c)$, the KL divergence term denotes the
divergence between two Bernoulli random variables with parameters
$\frac{p(m)}{p(S_m)}$ and $q_m(m)$ respectively. The last inequality follows from the
non-negativity of KL-divergence and because
\[
 \sup_{p\in\cP}\biggl[\sum_{x\geq m+1}p(x)\log\frac{p(x)}{q'(x)}+p(S_{m+1})\log\frac1{p(S_{m+1})}\biggr]
  \ge \tilde{\cT}_{m+1}(\cP). \eqed
  \]
\eLemmap

\bCorollary
\label{corr:tm}
For all collections $\cP\in\pn$, $\lim_{m\to\infty} \tilde{\cT}_m$ exists.
\Proof From Lemma~\ref{lm:tmni} and the fact that $\cT_m\ge 0$ for all $m$.
\eCorollary

\bLemma
\label{lm:tmt}
For all collections $\cP\in\pn$, the limit $\lim_{m\to\infty} \cT_m(\cP)$ exists and hence 
\[
  \cT(\cP)=\lim_{m\to\infty} \cT_m(\cP).
  \]
\Proof
If $\cP$ is not tight, the lemma holds vacuously from Proposition~\ref{prop:tight}.
Therefore, we suppose in the rest of the proof that $\cP$ is tight.
Observe from the definitions that 
\[
\cT_m(\cP)\leq \tilde{\cT}_m(\cP).
\]
Let $S_m=\sets{x\ge m}$ as before and let $q$ be any distribution over $\naturals$.
Then
\begin{align*}
&\sup_{p\in\cP}\sum_{x\geq m} p(x)\log\frac{p(x)}{q(x)}\\
 &=\sup_{p\in\cP} \bigg (\sum_{x\ge m} p(x)\log\frac{p(x)}{q(x)} +p(S_m)\log \frac{p(S_m)}{p(S_m)}\bigg)\\
 &\geq \sup_{p\in\cP} \bigg (\sum_{x\ge m} p(x)\log\frac{p(x)}{q(x)} + p(S_m)\log \frac1{p(S_m)} +\inf_{\hat{p}\in\cP}\hat{p}(S_m)\log \hat{p}(S_m)\bigg)\\
 &\geq\inf_{q'\in\pn} \sup_{p\in\cP} \bigg (\sum_{x\ge m} p(x)\log\frac{p(x)}{q'(x)}
+ p(S_m)\log \frac1{p(S_m)}\bigg) + \inf_{\hat{p}\in\cP}\hat{p}(S_m)\log \hat{p}(S_m)\\
 &= \tilde{\cT}_m(\cP)+ \inf_{\hat{p}\in\cP}\hat{p}(S_m)\log \hat{p}(S_m)
 \end{align*}
Since $\cP$ is tight, $\sup_{p\in\cP}p(S_m)\to0$ as $m\to\infty$ and hence $\inf_{\hat{p}\in\cP}\hat{p}(S_m)\log \hat{p}(S_m) \to0$ as 
$m\to\infty$. From Corollary~\ref{corr:tm}, we know that the sequence $\sets{\tilde{\cT}_m(\cP)}$ has a limit. Therefore, the sequence $\cT_m(\cP)$ also has a 
limit and in particular we conclude
\[
\tilde{\cT} =\lim_{m\to\infty} \cT_m(\cP).\eqed
\]
\eLemmap
Therefore, taking into account the above lemma, we can rephrase the definition
of tail redundancy as in~\eqref{dfn:tr},
\[
\cT(\cP)\ed \lim_{m\to\infty}\inf_{q_m\in\pn} \sup_{p\in\cP} \sum_{x\ge m} p(x) \log\frac{p(x)}{q_m(x)}
\]
We now show that 
\[
\cT(\cP)=\min_{q\in\pn} \lim_{m\to\infty}\sup_{p\in\cP} \sum_{x\ge m} p(x) \log\frac{p(x)}{q(x)}.
 \]
Note that the limit above need not be finite for every $q$. We take the above 
equation to mean the minimization over all $q$ such that the limit exists. 
If no such $q$ exists, the term on the right is considered to be vacuously 
infinite.

\bLemma
\label{lemaExs}
For a collection $\cP$ of distributions over $\naturals$ with tail redundancy $\cT(\cP)$, there is a distribution $q^*$ over $\naturals$ that satisfies
\[
 \lim_{m\to\infty} \sup_{p\in\cP} \sum_{x\ge m}p(x) \log \frac{p(x)}{q^*(x)} = \cT(\cP)
\]
\Proof If $\cP$ is not tight, the lemma is vacuously true and any $q$ is a ``minimizer''.

Therefore, we suppose in the rest of the proof that $\cP$ is tight.
From Lemma~\ref{LemTightness}, we can pick a finite number $m_r$ such
that
\[
\sup_{p\in\cP} p(x \ge m_r) \le \frac1{2^r},
\]
and let $q_r$ be any distribution that satisfies
\[
\sup_{p\in\cP} \sum_{x\ge m_r}p(x) \log \frac{p(x)}{q_r(x)}
\le\inf_q \sup_{p\in\cP} \sum_{x\ge m_r}p(x) \log \frac{p(x)}{q(x)} + \frac1r
= \cT_{m_r}(\cP) + \frac1r.
\] 
Since the limit of $\cT_{m}(\cP)$ as $m\to\infty$ is $\cT(\cP)$, we have
\[
 \lim_{r\to\infty} \sup_{p\in\cP} \sum_{x\ge m_r} p(x) \log \frac{p(x)}{q_r(x)}=\cT(\cP).
\]
Take 
\[
q^*(x) =\sum_{r\ge1} \frac{q_r(x)}{r(r+1)}.
\]
Now we also have for $r\ge2$ and any $m_r< m < m_{r+1}$ that
\begin{align*}
\sup_{p\in\cP} \sum_{x\ge m_r} p(x) \log \frac{p(x)}{q^*(x)}
&\ge 
\sup_{p\in\cP}
\Paren{p(m_r\le x < m )\log \frac{p(m_r\le x < m )}{q^*(m_r\le x < m )}
+
 \sum_{x\ge m} p(x) \log \frac{p(x)}{q^*(x)}}\\
&\ge 
\sup_{p\in\cP}
\Paren{p(m_r\le x < m )\log p(m_r\le x < m )
+
 \sum_{x\ge m} p(x) \log \frac{p(x)}{q^*(x)}}\\
&\ge 
-\frac{r}{2^r} +
\sup_{p\in\cP} \sum_{x\ge m} p(x) \log \frac{p(x)}{q^*(x)},
\end{align*}
where the first inequality is the logsum inequality and the last
inequality follows because $p(m_r \le x < m) \le p(m_r \le x) \le \frac1{2^r}<\frac1e$ for $r \ge 2$. Similarly, for $r\ge 2$ and $m_r< m < m_{r+1}$, we have
\begin{align*}
\sup_{p\in\cP} \sum_{x\ge m} p(x) \log \frac{p(x)}{q^*(x)}
&\ge
\sup_{p\in\cP}
\Paren{p(m \le x < m_{r+1} )\log \frac{p(m\le x < m_{r+1} )}{q^*(m\le x < m_{r+1} )}
+
 \sum_{x\ge m_{r+1}} p(x) \log \frac{p(x)}{q^*(x)}}\\
&\ge 
\sup_{p\in\cP}
\Paren{p(m \le x < m_{r+1} )\log p(m \le x < m_{r+1} )
+
 \sum_{x\ge m_{r+1}} p(x) \log \frac{p(x)}{q^*(x)}}\\
&\ge 
-\frac{r}{2^{r}} +
\sup_{p\in\cP} \sum_{x\ge m_{r+1}} p(x) \log \frac{p(x)}{q^*(x)}.
\end{align*}
\ignore{Therefore,
\[
\sup_{p\in\cP} \sum_{x\ge m_r} p(x) \log \frac{p(x)}{q^*(x)}
+\frac{r}{2^r} \ge
\sup_{p\in\cP} \sum_{x\ge m} p(x) \log \frac{p(x)}{q^*(x)}
\ge -\frac{r+1}{2^{r+1}} +
\sup_{p\in\cP} \sum_{x\ge m_{r+1}} p(x) \log \frac{p(x)}{q^*(x)}.
\]}
Therefore, 
\begin{align*}
&\limsup_{m\to\infty}\sup_{p\in\cP} \sum_{x\ge m} p(x) \log \frac{p(x)}{q^*(x)}\\
&\le
\lim_{r\to\infty} \left[ \sup_{p\in\cP} \sum_{x\ge m_r} p(x)\log \frac{p(x)}{q_r(x)} +\frac{r}{2^r} + \frac{\log r(r+1)}{2^r} \right] \\
&= \cT(\cP).
\end{align*}
Similarly, 
\begin{align*}
\liminf_{m\to\infty}\sup_{p\in\cP} \sum_{x\ge m} p(x) \log \frac{p(x)}{q^*(x)}
\ge \cT(\cP),
\end{align*}
and the lemma follows.
\eLemma

Henceforth, we will describe any distribution $q^*$ as in the lemma above
as ``$q$ achieves the tail redundancy for $\cP$''.

\bCorollary
\label{CorExs}
If a collection $\cP$ of distributions is tight and has tail redundancy $\cT(\cP)$, then there is a distribution $q^*$ over $\naturals$ that satisfies
\[
 \lim_{m\to\infty} \sup_{p\in\cP} \sum_{x\ge m}p(x) \log \frac{p(x)/\tau_p}{q^*(x)} = \cT(\cP)
\]
\Proof
The result follows using Lemma~\ref{lemaExs} and the fact that $\cP$ is tight. 
\eCorollary

\subsection{Properties of the tail redundancy}
\label{s:prop}
We examine two properties of tail redundancy in this subsection. Note
that the tail redundancy $\cT(\cP)$ is defined as the limit of
$\cT_m(\cP)$ as $m\to\infty$, however $\cT_m(\cP)$ need not always be
non-negative. However, we show that $\cT(\cP)$ is always non-negative.
The second property concerns the behavior of tail redundancy across
finite unions of classes. This property, while interesting inherently,
also helps us cleanly characterize the per-symbol redundancy of \iid
sources in Section~\ref{s:main}.

\bLemma For all $\cP$, $\cT(\cP)\geq 0$.  \Proof Again, if $\cP$ is
not tight, the lemma is trivially true from
Proposition~\ref{prop:tight}. Consider therefore the case where $\cP$
is tight. Fix $m\in\naturals$, and as before, let
$S_m=\{x \in\naturals:x\ge m\}$. Then,
\[
\inf_{q} \sup_{p\in\cP}\sum_{x\ge m}p(x)\log \frac{p(x)}{q(x)}
\geq \sup_{p\in\cP}   \bigg\{p(S_m)\log p(S_m)\bigg\}
\]
Furthermore, $\sup_p p(S_m)\log p(S_m)\to 0$ as $m\to\infty$ because
$\sup_p p(S_m)\to 0$ as $m\to\infty$. To see this, note that if $\sup_p p(S_m)\le \frac1e$, then 
\[
\sup_p p(S_m)\log p(S_m)\ge \Paren{\sup_{p'}p'(S_m)} \log 
\Paren{\sup_{p'}p'(S_m)}. 
\]
The lemma follows.
\ignore{Since $\cP$ is tight, $\sup p(S_m)$ goes to zero as $m\to\infty$. Note that for small value of $p(S_m)$,  $p(S_m)\log p(S_m)$ is a decreasing function of $p(S_m)$, so 
\[
\sup \left\{p(S_m)\log p(S_m)\right\}\geq \sup p(S_m)\log \sup p(S_m).
\]
Since $\sup p(S_m)\log \sup p(S_m)$ is a continuous function of $\sup p(S_m)$, it goes to $0$, as $\sup p(S_m)$ goes to $0$. Also $\sup \left\{p(S_m)\log p(S_m)\right\}\leq 0$ always holds, so $\lim_{m\to\infty}\sup \left\{p(S_m)\log p(S_m)\right\}$ goes to $0$ as $m$ goes to $\infty$. Therefore,
\[
\cT(\cP)=\lim_{m\to\infty}\inf_{q_m} \sup_{p\in\cP} \sum_{x>m}p(x)\log \frac{p(x)}{q(x)}\geq \lim \sup_{p\in\cP}   \left\{p(S_m)\log p(S_m)\right\}=0
\]}
\eLemma

We now show that the tail redundancy of a finite union of collections equals
the maximum of the tail redundancies of the individual parts of the union.
\bLemma
\label{lm:max}
Let $\cT(\cP_1),\cT(\cP_2),\dots, \cT(\cP_k)$ be tail redundancy of collections $\cP_1,\cP_2,\dots, \cP_k$ 
respectively. Then 
\[
\cT(\cup_{i=1}^k \cP_i)=\max_{1\leq j\leq k} \cT(\cP_j).
\]
\Proof
We first observe $\cT(\cup_{i=1}^k \cP_i)\geq \max_{1\leq j\leq k} \cT(\cP_j)$,
since for all $q\in\cP(\naturals)$, and all $1\le j\le k$, we have
\begin{align*}
\sup_{p\in\cup_i{\cP_i}} \sum_{x\ge m}p(x)\log \frac{p(x)}{q(x)} 
&\geq \sup_{p\in \cP_j} \sum_{x\ge m}p(x)\log \frac{p(x)}{q(x)}\\
&\geq \inf_{q'} \sup_{p\in \cP_j} \sum_{x\ge m}p(x)\log \frac{p(x)}{q'(x)}\\
&= \cT_m(\cP_j)
\end{align*}
\newcommand{\oq}{\hat{q}}
To show that $\cT(\cup_{i=1}^k \cP_j)\le \max_j \cT(\cP_j)$, 
let $q_1,q_2,\dots,q_k$ be distributions that achieve the tail redundancies $\cT(\cP_1),\cT(\cP_2),\dots,\cT(\cP_k)$ respectively.
Furthermore, for all distributions $q\in\cP(\naturals)$ and collections of
distributions $\cP\subset\cP(\naturals)$, let 
\[
\cT_m(\cP,q)=\sup_{p\in\cP} \sum_{x\ge m}p(x)\log \frac{p(x)}{q(x)}.
\]
Clearly, we have
\[
\cT(\cup_{i=1}^k \cP_i) =\lim_{m\to\infty}\inf_q \cT_m(\cup_{i=1}^k \cP_i,q).
\]
Let
\[
  \oq(x)=\frac{\sum_{i=1}^k q_i(x)}{k}
\]for all $x\in\naturals$.  We will
attempt to understand the behavior of the sequence
$\cT_m(\cup_{i=1}^k \cP_i,\oq)$ first.  Observe that
\begin{align}
\nonumber \cT_m(\cup_{i=1}^k \cP_i,\oq)
&= \max_{1\leq j\leq k} \sup_{p\in\cP_j}\sum_{x\geq m}p(x)\log \frac{p(x)}{\oq(x)}\\
&=\max_{1\leq j \le k} \cT_m(\cP_j,\oq).\label{eq:tu}
\end{align}
For all $1\le j \le k$, the limit $\lim_{m\to\infty} \cT_m(\cP_j,\oq)$ exists and is equal to $\cT(\cP_j)$. This follows because for all $j$,
\begin{align*}
\cT_m(\cP_j,\oq)
&\ale{(a)}\sup_{p\in\cP_j}\left (\sum_{x\geq m}p(x)\log \frac{p(x)}{q_j(x)} + \sum_{x\geq m}p(x) \log k\right )\\
&\leq \sup_{p\in\cP_j}\left (\sum_{x\geq m}p(x)\log \frac{p(x)}{q_j(x)}\right) 
+\sup_{p\in\cP_j} \sum_{x\ge m} p(x) \log k,
\end{align*}
where $(a)$ follows because for all $x$, $q(x)\ge q_j(x)/k$.
Let $\delta_{m,j}=\sup_{p\in\cP_j}\sum_{x\ge m} p (x)$. Note that since $\cP_j$ is tight, we have
$\lim_{m\to\infty} \delta_{m,j}=0$.
Thus
\[
\inf_q\sup_{p\in\cP_j}\sum_{x\geq m}p(x)\log \frac{p(x)}{q(x)}
\le 
\cT_m(\cP_j,\oq)
\le
\sup_{p\in\cP_j}\left (\sum_{x\geq m}p(x)\log \frac{p(x)}{q_1(x)}\right) + \delta_{m,j} \log k 
\]
and both the lower and upper bound on $\cT_m(\cP_j,\oq)$, $m\ge 1$, are sequences whose limit exists, and both limits are $\cT(\cP_j)$ as $m\to\infty$.
We conclude then, that for all $1\le j \le k$,
\begin{equation}
\label{eq:limtpj}
\lim_{m\to\infty}\cT_m(\cP_j,\oq) = \cT(\cP_j).
\end{equation}

Recalling from~\eqref{eq:tu} that
$\cT_m(\cup_{i=1}^k \cP_i,\oq) = \max_i \cT_m(\cP_i,\oq)$,
using Equation~\eqref{eq:limtpj} and Lemma~\ref{seqLemma} that shows
that the limit of the maximum of a finite number of sequences equals
the maximum of their limits, we have that the sequence
$\cT_m(\cup_{i=1}^k \cP_i,\oq)$, $m\ge1$, also has a limit and that
\[
\lim_{m\to\infty} \cT_m(\cup_{i=1}^k \cP_i,\oq) = \max_{1\le j\le k} \cT(\cP_j).
\]
Putting it all together, we have
\[
\cT(\cup_{i=1}^k \cP_i) =\lim_{m\to\infty}\inf_q \cT_m(\cup_{i=1}^k \cP_i,q)
\le\lim_{m\to\infty} \cT_m(\cup_{i=1}^k \cP_i,\oq) = \max_{1\le j\le k} \cT(\cP_j).\]
The lemma follows.
\eLemma

\section{Main result}
\label{s:main}
In \cite{HS14} we showed that if a collection of distributions has
finite single letter redundancy, then a couple of technical
conditions, one of which was similar to but not the same as the the
tail redundancy condition being 0, then the collection was strongly
compressible. At the same time, we also had noted that the technical
conditions therein were not necessary. The main result of this section
is that the per-symbol redundancy goes to tail redundancy as $n$
increases. In fact, we first show that per-symbol redundancy is always
greater than or equal to tail redundancy as $n\to\infty$ and
conversely we show that tail redundancy is always greater than or
equal per-symbol redundancy as $n\to\infty$. This result implies that
zero tail redundancy is a necessary and sufficient condition for a
collection to be strongly compressible.

\bLemma
\label{infLemma}
For any $\cP\subset \prob(\naturals)$, if $R_1(\cP)=\infty$ then $\cT(\cP)=\infty$. 
\Proof
Assume on the contrary that $\cT(\cP)$ is finite, then for some $m$, we will have
\[
\inf_q \sup_p \sum_{x\ge m} p(x)\log \frac{p(x)}{q(x)}=M<\infty.
\]
Fix $\epsilon>0$ and let $q_m\in\prob(\naturals)$ be any distribution that incurs tail redundancy $\le M+\epsilon$, i.e.
\[
\sup_p \sum_{x\ge m}p(x)\log \frac{p(x)}{q_m(x)}\le M+\epsilon.
\]
Consider $q_0=(\frac1{m},\dots,\frac1{m})$, a uniform distribution over $\{1,2,3,\dots,m\}$, and let $q(x)=\frac{q_0(x)+q_m(x)}{2}$ for all $x\in\naturals$. Then for all $p\in\cP$,
\[
R_1=\sum_{x=1}^{\infty} p(x)\log \frac{p(x)}{q(x)}\leq \log 2m+M+\epsilon+1<\infty,
\]
which is a contradiction.
\eLemma

\bTheorem
\label{Thm1}
Let $\cP$ be a collection of distributions over $\naturals$ and
$\cP^\infty$ be the collection of all measures over infinite sequences
that can be obtained by \iid sampling from a distribution in $\cP$. 
Then
\[
R(\cP^\infty) = \limsup_{n\to \infty}  \frac1n R_n(\cP^\infty)=\cT(\cP).
\]
\eTheorem

A couple of quick examples first.

\bExample Proposition~\ref{prop2} proved that $\cB^\infty$ does not
have zero asymptotic per-symbol redundancy, and we note that
$\cT(\cB)=1$.  \eExample

\ignore{\bExample
\label{exm3}
Let $h>0$. Let $\mathcal{M}_h$ be the collection of uniform
distributions over $\naturals$ such that
\[
E_p \bigg(\log \frac1{p(X)}\bigg)^2<h.
\]
Let $\mathcal{M}_h^\infty$ be the set of all \iid distributions with
one dimensional marginals from $\mathcal{M}_h$. Then it is easy to
verify that $\cT(\cM_h)=0$ and that $\mathcal{M}_h^\infty$ is strongly
compressible. Specifically, we can construct a measure $q^*$ over infinite
sequences of naturals whose per-symbol length-$n$ redundancy against 
sources in $\cM_h^\infty$ is upper bounded by (see~\cite{SAKSisit14})
\[
\frac{2h^{3/2}}{\sqrt {\log {n}}}+\pi\sqrt{\frac{2}{3n}} 
\]
so $\mathcal{M}_h^\infty$ is strongly compressible.
\eExample}

\section{Proof of Theorem~\ref{Thm1}}
We first consider the case when $\cP$ is not tight. Using
Lemma~\ref{LemTightness}, $R_1=\infty$ and using Lemma~\ref{infLemma},
the tail redundancy $\cT(\cP)=\infty$ as well. Furthermore, $R_1=\infty$
implies from Proposition~\ref{prop:one} that for all $m\ge1$,
$\frac1m R_m(\cP^\infty)=\infty$. Therefore, if $\cP$ is not tight,
$\frac1m R_m(\cP^\infty)=\cT(\cP)=\infty$ and the Theorem holds.

We now consider the case when $\cP$ is tight.  If $\cP$ is tight we
first show in Section~\ref{s:direct} that for all $m$,
$\frac1m R_m(\cP^\infty)\geq \cT(\cP)$ and in Section~\ref{s:converse}
that $\frac1m R_m(\cP^\infty)\leq \cT(\cP)$.

\subsection{Direct part}
\label{s:direct}
%
We show that if $\cP$ is tight, for all $n\ge 1$,
\[
  \frac1n R_n(\cP^\infty)\geq \cT(\cP),
\]
thus also proving that $R(\cP^\infty)\ge \cT(\cP)$.
Since $\cP$ is tight, for any $c>0$, we can find a finite number $m_n$ such that
$\forall p\in\cP$,
\[
p(x\ge m_n)<\frac{c}{n}.
\]
Let $\tau_n^p \ed p(x\ge m_n)$ be the tail probability
past $m_n$ under $p$.
Let $\cY=\{-1,1,2,\dots,m_n-1\}$.
For each sequence $x^n\in\naturals^n$, let the auxiliary sequence $y^n= y_1\upto y_n\in\cY^n$ be defined
by
\[
y_i(x^n)=\left\{ 
  \begin{array}{l l}
   x_i  & \quad  \text{if}\quad x_i< m_n\\
    -1  & \quad  \text{if}\quad x_i\ge m_n.
  \end{array} \right.
\]
For $x^n\in\naturals^n$ and $y^n\in\cY^n$, we say $x^n\sim y^n$ if $x^n$ and
$y^n$ are consistent ($y^n$ would be the auxiliary sequence constructed from
$x^n$).
For all $n\geq 1$ and $r_n\in\pnn$, we show that
\[
\frac1n\sum p(x^n)\log\frac{p(x^n)}{r_n(x^n)}\geq \cT(\cP).
\]
proving also that
\[
R(\cP^\infty)= \limsup_{n\to\infty} \frac1n R_n(\cP^\infty) = \limsup_{n\to\infty}\inf_{r_n\in\pnn} \sup_{p\in\cP^\infty}\frac1n\sum p(x^n)\log\frac{p(x^n)}{r_n(x^n)}\geq \cT(\cP).
\]
%
Fix any $r_n\in\pnn$. Now for all $x^n\in\naturals^n$ and $y^n$ such that
$x^n\sim y^n$,
\[
r_n(x^n)=r_{XY}(x^n,y^n)=r_{X|Y}(x^n|y^n)r_Y(y^n),
\]
where we use the subscripts $XY$, $X|Y$ and $Y$ to denote the
appropriate induced distributions, \ie
$r_Y(y^n) =\sum_{x^n\sim y^n} r_n(x^n)$, $r_{XY}(x^n, y^n) = r_n(x^n)$
if $x^n\sim y^n$ and 0 else, and
$r_{X|Y}(x^n|y^n) = r_n(x^n)/r_{Y}(y^n)$ if $x^n\sim y^n$ and 0 else.
Define $G \subset\cY^n$, where
\[
  G=\{y^n\in\cY^n:\text{exactly one element of } y^n \text{ is} -1\}.
\]
We will focus our attention primarily on auxiliary sequences in $G$, by noting
\begin{align*}
&\sup_{p\in\cP^n} D_n(p||r_n)
=\sup_{p\in \cP^n} \sum_{x^n} p(x^n)\log\frac{p(x^n)}{r_n(x^n)}\\
&=\sup_{p\in\cP^n}\bigg(\sum_{y^n\in\cY^n}p(y^n)\sum_{x^n}p(x^n|y^n)\log\frac{p(x^n|y^n)}{r_{X|Y}(x^n|y^n)}
+ \sum_{y^n\in\cY^n} p(y^n)\sum_{x^n}p(x^n|y^n)\log\frac{p(y^n)}{r_Y(y^n)}\bigg)\\
&\overset{(a)}{\geq}\sup_{p\in\cP^n} \sum_{y^n\in G} p(y^n)\sum_{x^n\sim y^n}
 p(x^n|y^n)\log\frac{p(x^n|y^n)}{r_{X|Y}(x^n|y^n)},
\end{align*}
where $(a)$ uses the fact that KL divergence is greater than or equal
to 0, and in the above, we use the convention that if
$x^n\not\sim y^n$, both $p(x^n|y^n)$ and $r_n(x^n|y^n)$ are 0 and the
corresponding contribution of such a pair to the summations above is
0.

For a given consistent pair $x^n\sim y^n$ for $y^n\in G$, it is
easy to see that if $y_i=-1$, then $x_i$ is the only symbol $\ge m_n$
and $p(x^n|y^n)$ is essentially written in terms of the single letter
distribution on $\naturals$ which we denote $p(\cdot|X\ge m_n)$, namely
\[
  p(x_i|X\ge m_n)\ed p(x^n|y^n)=\frac{p(x_i)}{\sum_{x'\ge m_n} p(x')} = \frac{p(x_i)}{\tau^p_n}
\]
Similarly, for any $y^n\in G$, we can extract a single letter distribution
$\ryn\in \pn$ from $r_{X|Y}(x^n|y^n)$ in a similar
fashion. If $i$ is the only number such that $y_i=-1$, then for all
$x^n \sim y^n$ 
\[
\ryn(x_i) \ed r_{X|Y}(x^n|y^n) = \frac{r_n(y_1^{i-1}x_i y_{i+1}^n)}{\sum_{x\ge m_n} r_n(y_1^{i-1}x y_{i+1}^n)} 
\]
Therefore,
\begin{align*}
&\sup_{p\in\cP^n} \sum_{y^n\in G} p(y^n)\sum_{x^n\sim y^n}
p(x^n|y^n)\log\frac{p(x^n|y^n)}{r_n(x^n|y^n)}\\
&\geq\sup_{p\in\cP}\sum_{y^n\in G}p(y^n)\sum_{x\ge m_n} p(x|X\ge m_n)\log\frac{p(x|X\ge m_n)}{\ryn(x)}\\
&=\sup_{p\in\cP} \sum_{y^n\in G}\frac{p(y^n)}{\tau^p_n}\sum_{x\ge m_n}p(x)\log \frac{p(x)/\tau^p_n}{\ryn(x)}.
\end{align*}
To reduce the above expression, for all $p\in\cP$, let
\[
\y(p)=\arg \min_{y^n\in G} \sum_{x\geq m_n}p(x)\log\frac{p(x)}{\ryn(x)}.
\]
Then,
\[
\sup_p \sum_{y^n\in G}\frac{p(y^n)}{\tau^p_n}\sum_{x>	m_n}p(x)\log \frac{p(x)/\tau^p_n}{\ryn(x)}
\geq 
\sup_p \frac{p(G)}{\tau^p_n}\sum_{x\ge m_n} p(x) \log \frac{p(x)/\tau^p_n}{\ryp(x)}.
\]
Observing that for all $p\in\cP$, $p(G)=n(1-\tau_n^p)^{n-1}\tau_n^p$, we have
therefore that
\begin{align}
\label{eqDirect1}
\sup_p D_n(p_{X^n}||r_{X^n})\nonumber
&\geq \sup_p \frac{p(G)}{\tau_n^p}\sum_{x\ge m_n}p(x)\log\frac{p(x)/\tau_p^n}{\ryp(x)}\\\nonumber
&= \sup_p \frac{p(G)}{\tau_n^p}\bigg(\sum_{x\ge m_n} p(x)\log\frac{p(x)}{\ryp(x)}+\tau^p_n\log\frac1{\tau^p_n}\bigg)\\
&\geq \sup_p n\Paren{1-\frac{c}{n}}^n
\bigg(\sum_{x\ge m_n} p(x)\log\frac{p(x)}{\ryp(x)}+\tau^p_n\log\frac1{\tau^p_n}\bigg).
\end{align}
For $\y\in G$, let $\cP_\y=\{p\in\cP: \y(p) =\y\}$. Then $\cP$ can be written
as the finite union,
\[\cP=\cup_{\y\in G} \cP_{\y}.
\]
Therefore from Lemma~\ref{lm:max}
\[
\cT(\cP)=\max_{\y\in G}  \cT(\cP_{\y}).
\]
We then have
\begin{align}
\label{eqDirect2}
\sup_{p\in\cP} \left[\sum_{x\ge m_n}p(x)\log \frac{p(x)}{\ryp(x)}+\tau^{p}_n\log\frac1{\tau^{p}_n}\right]\nonumber
&=\max_{\y\in G} \sup_{p\in\cP_{\y}}\left[\sum_{x\ge m_n}p(x)\log \frac{p(x)}{r^\y(x)}+\tau^{p}_n\log\frac1{\tau^{p}_n}\right]\\\nonumber
&\geq \max_{\y\in G}\left(\inf_{q_{\y}\in\pn}\sup_{p\in\cP_{\y}}\left[\sum_{x\ge m_n}p(x)\log \frac{p(x)}{q_{\y}(x)}+\tau^{p}_n\log\frac1{\tau^{p}_n}\right]\right)\\\nonumber
&= \max_{\y \in G} \tilde{\cT}_{m_n} (\cP_{\y})\\\nonumber
&\overset{(*)}{\geq} \max_{\y\in G} \cT(\cP_{\y})\\
&=\cT(\cP),
\end{align}
where $(*)$ follows since for any collection, Lemmas~\ref{lm:tmni} and~\ref{lm:tmt} together imply that $\tilde{\cT}_m$ monotonically decreases to the limit $\cT$. 
Putting~\eqref{eqDirect1} and~\eqref{eqDirect2} together, we obtain
\[
\sup_{p\in\cP} D_n(p||r)\geq n\Paren{1-\frac{c}{n}}^n\cT(\cP).
\]
Since the inequality holds for all $c>0$, we have
\[
\sup_{p\in\cP}\frac1{n}D_n(p||r)\geq \sup_{c>0} \Paren{1-\frac{c}{n}}^n\cT(\cP) = \cT(\cP).
\] 
\subsection{Converse part}
\label{s:converse}
We now show that
\[
R(\cP^\infty)= \limsup_{n\to\infty}\frac1{n}R_n(\cP^\infty)\leq \cT(\cP).
\]
First, note from Proposition~\ref{prop:one} that 
\[
R(\cP^\infty)=  \limsup_{n\to\infty}\frac1{n}R_n(\cP^\infty) \le R_1(\cP) 
\]
so if $R(\cP^\infty)$ is $\infty$, so is
$R_1(\cP)$, and from Proposition~\ref{prop:tight}, $\cT(\cP)$ is
infinite as well, and vacuously, $R(\cP^\infty) \le \cT(\cP)$.

For the rest of the proof, we assume that 
$R(\cP^\infty)< \infty$.



Our proof will be constructive. We describe length $n$ sequences from
$\naturals^n$ using distributions $q_n\in\pnn$, constructed as
follows. We first clip the sequences at a threshold $m$, replacing all
occurrences of numbers $\ge m$ in the sequence with a new symbol,
-1. To complete the description, we then describe the actual number
that occurred corresponding to each -1 using a single letter
distribution $q^*\in\pn$ that achieves the tail redundancy of
$\cP$. The threshold $m$ will be chosen to vary with the sequence
length $n$. This simple construction is enough to achieve
asymptotically per-symbol redundancy of $\le \cT(\cP)$.

While this approach will yield a sequence of distributions
$\sets{q_n\in\pnn, n\ge 1}$ that are not consistent (primarily because
we vary the threshold $m$ with the sequence length $n$), note that the
general construction in Appendix~\ref{app:rdetails} provides a way to
construct a universal probability measure $q\in\pni$ which incurs the
same asymptotic per-symbol redundancy as the sequence
$\sets{q_n\in\pnn, n\ge 1}$ of distributions. 

We begin by noting that for any finite $m$, there is a distribution
$r_n$ that achieves the minimax redundancy of encoding $m-$ary \iid
strings~\cite{XB97}.\footnote{Such a $r_n$ is also a Bayesian mixture
  of the $m-$ary \iid probability measures.}  Let the redundancy of
$r_n$ against $(m+1)-$ary \iid sequences of length $n$ be
$\rho_{m,n}$. It is known that
$\rho_{m,n}\sim \frac{m}2\log n$~\cite{dav73,dav81,xb00}.  In
particular, it is easy to see that encoding these sequences with the
(suboptimal) add-1 (Laplace estimator, or the Bayesian mixture with
the conjugate Dirichlet prior with all parameters 1) rule incurs
redundancy $\log{n+m-1 \choose m-1}$, so
\begin{equation}
  \label{eq:laplace}
  \rho_{m,n} \le \log{n+m-1 \choose m-1} \le (m-1)\log n.
\end{equation}

Let $q^*$ be the distribution that achieves $\cT(\cP)$. With the
benefit of hindsight, we set $m=\sqrt{n}$. As before, let
$\cY= \sets{-1,1\upto m-1}$ and we construct an auxiliary sequence
$y^n\in\cY^n$ from $x^n$ where
\[
y_i=\left\{ 
  \begin{array}{l l}
   x_i  & \quad  \text{if}\quad x_i< m\\
    -1  & \quad  \text{if}\quad x_i\ge m.
  \end{array} \right.
\]
As before, given any sequence $y^n\in\cY^n$, and $x^n\in\naturals^n$,
we say $x^n\sim y^n$ if $y^n$ is consistent with $x^n$ ($y^n$ would be
constructed from $x^n$).  Let
$q_m(x) = q^*(x)/\sum_{x'\geq m} q^*(x')$ for $x\ge m$.  Then, we
construct a distribution $q_n\in\pnn$ by first specifying the probabilities
of the auxiliary sequences $y^n\in \cY^n$ using the $m-$ary minimax optimal
distribution $r_n$,
\[
q_Y(y^n)= r_n(y^n),
\]
followed by describing $x_i$ for each $y_i$,
\[
q_{X|Y}(x^n|y^n)=\prod_{i=1}^n q(x_i|y_i),
\]
where
\[
q(x_i|y_i)= \left\{ 
  \begin{array}{l l}
 q_m(x_i)  & \quad  \text{if}\quad y_i=-1\\
    1  & \quad \text{ if } y_i\ne -1, x_i=y_i\\
    0  & \quad \text{ if } y_i\ne -1, x_i\ne y_i.\
  \end{array} \right.\]
Finally for all $x^n$, $q(x^n) = \sum_{z^n\in\cY^n} q_Y(z^n) q_{X|Y}(x^n|z^n)$,
which will coincide with $q_Y(y^n)q_{X|Y}(x^n|y^n)$ for the unique $y^n\in\cY^n$
that is consistent, \ie $x^n\sim y^n$.
Then,
\begin{align}
\label{eq1Con}
&\frac1{n}\sum_{x^n\in\naturals^n} p(x^n)\log\frac{p(x^n)}{q(x^n)}\nonumber\\
&=\frac1{n} \sum_{y^n\in\cY^n} p(y^n)\log\frac{p(y^n)}{q(y^n)}+  \frac1{n}\sum_{\substack{x^n\in\naturals^n, y^n\in\cY^n\\x^n\sim y^n}} p(x^n)\log\frac{p(x^n|y^n)}{q(x^n|y^n)}\nonumber\\
&= \frac{\rho_{m,n}}{n}+\frac1{n}\sum_{y^n\in\cY^n} p(y^n)\sum_{\substack{x^n\in\naturals^n\\x^n\sim y^n}}p(x^n|y^n)\log\frac{p(x^n|y^n)}{q(x^n|y^n)}.
\end{align}

For any $y^n\in\cY^n$, let $k(y^n)$ be the number of occurrences of $-1$ in $y^n$. Let $\tau_{p,m}=\sum_{x\ge m} p(x)$, then for all $x^n\sim y^n$, 
\[
p(x^n|y^n)=\prod_{i:y_i=-1}\frac{p(x_i)}{\tau_{p,m}}.
\]
We can rewrite the second term in equation~\eqref{eq1Con} as 
\begin{align*}
\frac1{n}\sum_{y^n\in\cY^n} p(y^n)&\sum_{\substack{x^n\in\naturals^n\\ x^n\sim y^n}}p(x^n|y^n)\log\frac{p(x^n|y^n)}{q(x^n|y^n)}\\
&=\frac1{n}\sum_{y^n\in\naturals^m} p(y^n)
\sum_{\substack{x^n\in\naturals^n\\ x^n\sim y^n}}\Paren{\prod_{j:y_j=-1}\frac{p(x_j)}{\tau_{p,m}}}
\log\prod_{i:y_i=-1}\frac{p(x_i)/\tau_{p,m}}{q_m(x_i)}\\
&=\frac1{n}\sum p(y^n) A(k(y^n)).
\end{align*}
For each $y^n$, we can bound $A(k(y^n))$ as follows,
\[
\label{eqBound}
A(k(y^n))
=\sum_{j:y_j=-1}\sum_{x_j\geq m} \frac{p(x_j)}{\tau_{p,m}}\log\frac{p(x_j)/\tau_{p,m}}{q_m(x_j)} \\
= k(y^n) \sum_{x\geq m}\frac{p(x)}{\tau_{p,m}}\log\frac{p(x)/\tau_{p,m}}{q_m(x)}.
\]
Now we have,
\begin{align}
\label{numiny}
\sum_{y^n} p(y^n)A(k(y^n))
&= \sum_{y^n} p(y^n)k(y^n) \sum_{x\geq m}\frac{p(x)}{\tau_{p,m}}\log\frac{p(x)/\tau_{p,m}}{q_m(x)}\nonumber\\
&= \sum_{x\geq m}\frac{p(x)}{\tau_{p,m}}\log\frac{p(x)/\tau_{p,m}}{q_m(x)} \sum_{y^n} p(y^n)k(y^n)\nonumber\\
  &=\sum_{x\geq m}\frac{p(x)}{\tau_{p,m}}\log\frac{p(x)/\tau_{p,m}}{q_m(x)} {\mathbb E}(k(Y^n))
\end{align}
Combining equation~\eqref{eq1Con} and~\eqref{numiny}, we have
\begin{align*}
\frac1{n}\sum_{x^n} p(x^n)\log\frac{p(x^n)}{q(x^n)}
&\leq\frac{\rho_{m,n}}{n}+ \frac1{n}{\mathbb E}k(Y^n)\bigg(\sum_{x\geq m}\frac{p(x)}{\tau_{p,m}}\log\frac{p(x)/\tau_{p,m}}{q_m(x)}\bigg)\\
  &\leq\frac{\rho_{m,n}}{n}+\sum_{x\geq m}p(x)\log\frac{p(x)/\tau_{p,m}}{q_m(x)}\\
  &\leq\frac{\rho_{m,n}}{n}+\sum_{x\geq m}p(x)\log\frac{p(x)/\tau_{p,m}}{q^*(x)},
\end{align*}
where the second to last inequality follows since
${\mathbb E}k(Y^n)=n\tau_{p,m}$ and the last inequality because
$q_m(x)= q^*(x)/(\sum_{x'\ge m} q^*(x'))\ge q^*(x)$. Now, taking the supremum over all $p$ and the
limsup as $n\to\infty$, we have
\begin{align*}
  \limsup_{n\to\infty}&\sup_{p\in\cP^n}\frac1{n}\sum_{x^n\in\naturals^n} p(x^n)\log\frac{p(x^n)}{q(x^n)}\leq
  \\
  &\limsup_{n\to\infty} \sup_{p\in\cP}\left[\frac{\rho_{m,n}}{n}
+ \sum_{x\geq m}p(x)\log\frac{p(x)}{q^*(x)} + \tau_{p,m}\log\frac1{\tau_{p,m}}
\right].
\end{align*}
We claim that the limit above is $\cT(\cP)$. Now
from~\eqref{eq:laplace}, for all $m$ and $n$,
\[
  \frac{\rho_{m,n}}{n} \le \frac{m-1}n \log n,
\]
so as $n\to \infty$ while $m=\sqrt{n}$, $\frac{\rho_{m,n}}n\to 0$.
The second term in the parenthesis goes to $\cT(\cP)$ as $n\to\infty$,
since $m=\sqrt{n}\to\infty$ and since $q^*$ achieves the tail
redundancy for the collection $\cP$. For the last term, recall that
when the asymptotic per-symbol redundancy is finite, so is the single
letter redundancy. Therefore, the collection $\cP$ is tight, and hence
as $n\to \infty$, while $m=\sqrt{n}$,
\[
\lim_{n\to\infty}  \sup_{p\in\cP} \tau_{p,\sqrt{n}} \log \frac 1{\tau_{p,\sqrt{n}}} \to 0
\]
Therefore,
\[
\limsup_{n\to\infty}\frac1{n}R_n(\cP^\infty,q)\leq \cT(\cP).
\] 
\section{Conclusion}
The paper establishes the scaling of the asymptotic per-symbol redundancy,
and in particular shows that it is captured by the complexity in the tails
of the distributions of the class. As remarked earlier, one way to interpret
the result is to see the tail redundancy as the asymptotic cost of describing
novel symbols and the proofs of the lower and upper bounds bear out this
interpretation.

In future work, it will be useful to characterize this cost for
specific distribution classes, in particular for polynomial tail
distributions.  Note that distribution collections that scale
polynomially (Zipf-like $1/n^{1+\epsilon}$) can be constructed, and
these will have tail redundancy bounded away from 0 depending on how
these classes are constructed. Given the prevalence of these models in
language descriptions, this phenomenon naturally has implications in how
we interpret description of novel words.

It is also open at this point to characterize the equivalent of tail
redundancy for Markov sources over countably infinite
alphabets. Novelty here need not necessarily come from new symbols,
but also new states. The challenge here would be to restrict the classes
in a meaningful way that captures applications, yet provides insights
on encountering novelty from a average minimax redundancy perspective.

\section*{Acknowledgments}
This work was supported in part by the NSF Science \& Technology
Center for Science of Information Grant number CCF-0939370, as well as
NSF Grants CCF-1065632 and CCF-1619452.

%
\ignore{\section{Conclusion}
\label{SecConc}
We have obtained necessary and sufficient conditions on a family of
distributions to be strongly compressible, when the alphabet size is
countably infinite. We showed that if a family of distributions is
strongly compressible, then the single letter redundancy of very large
elements needs to go to zero. Conversely, we proved that zero single
letter redundancy of very large elements along with finite single
letter redundancy guarantee that a a family of distributions is
strongly universal compressible. While the conditions for
compressibility of a family of distribution over finite alphabet are
well known in literature, there are no similar results for infinite
alphabet in the case of strong compression. One further extension of
the paper could be obtaining the same results for a family of Markov
or stationary ergodic distributions.}

\bibliographystyle{unsrt}
\bibliography{univcod}

\begin{thebibliography}{10}

\bibitem{HS19:isit}
M.~Hosseini and N.~Santhanam.
\newblock Tail redundancy and its connections with universal compression.
\newblock Proceedings of IEEE Symposium on Information Theory, 2019.

\bibitem{dav73}
L.D. Davisson.
\newblock {Universal noiseless coding}.
\newblock {\em IEEE Transactions on Information Theory}, 19(6):783---795,
  November 1973.

\bibitem{Goo53}
I.J. Good.
\newblock The population frequencies of species and the estimation of
  population parameters.
\newblock {\em Biometrika}, 40(3/4):237---264, December 1953.

\bibitem{OSZ03:s+f}
A.~Orlitsky, N.P. Santhanam, and J.~Zhang.
\newblock Always {G}ood {T}uring: {A}symptotically optimal probability
  estimation.
\newblock {\em Science}, 302(5644):427---431, October~17 2003.
\newblock See also~{\it Proceedings of the}~44th~{\it Annual Symposium on
  Foundations of Computer Science},~October~2003.

\bibitem{mdlbook}
P.~Grunwald.
\newblock {\em {The Minimum Description Length Principle}}.
\newblock {MIT Press}, 2007.

\bibitem{risbook}
J.~Rissanen.
\newblock {\em Optimal Estimation of Parameters}.
\newblock {Cambridge University Press}, 2012.

\bibitem{CB90}
B.S. Clarke and A.R. Barron.
\newblock {Information theoretic asymptotics of Bayes methods}.
\newblock {\em itt}, 36(3):453---471, May 1990.

\bibitem{IH72}
I.~Ibragimov and R.~Hasminskii.
\newblock On the information in a sample about a parameter.
\newblock In {\em Proceedings of IEEE Symposium on Information Theory}, pages
  295--309, New York, 1972.

\bibitem{Gal79}
R.~Gallager.
\newblock Source coding with side information and universal coding.
\newblock Technical Report LIDS-P-937, Laboratory for Information and Decision
  Systems, MIT, 1979.

\bibitem{E80}
S.~Efroimovich.
\newblock Information contained in a sequence of observations.
\newblock {\em Problems of Information Transmission}, 15:178--179, 1980.

\bibitem{DL80}
L.~Davisson and A.~Leon-Garcia.
\newblock A source matching approach to finding minimax codes.
\newblock {\em IEEE Transactions on Information Theory}, 26(2):166--174, Mar
  1980.
\newblock 1980.

\bibitem{CB94}
B.S. Clarke and A.R. Barron.
\newblock {Jeffreys' prior is asymptotically least favorable under entropy
  risk}.
\newblock {\em Journal of Statistical Planning and Inference}, 41(1):37---60,
  1994.

\bibitem{XB97}
Q.~Xie and A.~Barron.
\newblock Minimax redundancy for the class of memoryless sources.
\newblock {\em IEEE Transactions on Information Theory}, 43(2):647---657, March
  1997.

\bibitem{xb00}
Q.~Xie and A.R. Barron.
\newblock Asymptotic minimax regret for data compression, gambling and
  prediction.
\newblock {\em IEEE Transactions on Information Theory}, 46(2):431---445, March
  2000.

\bibitem{BC88}
A.~Barron and T.~Cover.
\newblock A bound on the financial value of information.
\newblock {\em IEEE Transactions on Information Theory}, 34:1097--1100, 1988.

\bibitem{HO97}
David Haussler and Manfred Opper.
\newblock Mutual information, metric entropy and cumulative relative entropy
  risk.
\newblock {\em The Annals of Statistics}, 25(6):2451--2492, 1997.

\bibitem{os04:soi}
A.~Orlitsky and N.P. Santhanam.
\newblock Speaking of infinity.
\newblock {\em IEEE Transactions on Information Theory}, 50(10):2215---2230,
  October 2004.

\bibitem{wvk11}
A.~B. Wagner, P.~Viswanath, and S.~R. Kulkarni.
\newblock Probability estimation in the rare events regime.
\newblock {\em IEEE Transactions on Information Theory}, 57(6):3207--3229, Sep
  2011.

\bibitem{SW12}
W.~Szpankowski and M.~Weinberger.
\newblock Minimax pointwise redundancy for memoryless models over large
  alphabets.
\newblock {\em IEEE Transactions on Information Theory}, 58(7):4094--4104, Jul
  2012.

\bibitem{dav81}
M.~B. Pursley, M.~Wallace, L.D. Davisson, and R.~J. McEliece.
\newblock {Efficient universal noiseless source codes}.
\newblock {\em IEEE Transactions on Information Theory}, 279(3):269---279, May
  1981.

\bibitem{ds04}
M.~Drmota and W.~Szpankowski.
\newblock Precise minimax redundancy and regrets.
\newblock {\em IEEE Trans. Information Theory}, 50:2686--2707, 2004.

\bibitem{STW95}
Y.M. Shtarkov, T.J. Tjalkens, and F.M.J. Willems.
\newblock Multialphabet universal coding of memoryless sources.
\newblock {\em Problems of Information Transmission}, 31(2):114---127, 1995.

\bibitem{Wil98}
Frans M.~J. Willems.
\newblock The context-tree weighting method: Extensions.
\newblock {\em IEEE Transactions on Information Theory}, 44:792--798, 1998.

\bibitem{CsSh96}
I.~Csiszar and P.~Shields.
\newblock {Redundancy rates for renewal and other processes}.
\newblock {\em IEEE Transactions on Information theory}, 42:2065--2072, 1996.

\bibitem{FS02}
P.~Flajolet and W.~Szpankowski.
\newblock {Analytic variations on redundancy rates of renewal processes}.
\newblock {\em IEEE Transactions on Information theory}, 48:2911--2921, 2002.

\bibitem{WRF95}
J.~Rissanen M.~Feder M.~J.~Weinberger.
\newblock {A universal finite memory source}.
\newblock {\em IEEE Transactions on Information theory}, 41:643--652, 1995.

\bibitem{SAS22:jmlr}
N.~Santhanam, V.~Anantharam, and W.~Szpankowski.
\newblock Data driven weak universal compression.
\newblock {\em Journal of Machine Learning Research}, 23, 2022.

\bibitem{MF95}
N.~Merhav and M.~Feder.
\newblock A strong version of the redundancy capacity theorem.
\newblock {\em IEEE Transactions on Information Theory}, 41(3):714---722, May
  1995.

\bibitem{BGG08}
S.~Boucheron, A.~Garivier, and E.~Gassiat.
\newblock Coding on countably infinite alphabets.
\newblock Available from arXiv doc id: 0801.2456, 2008.

\bibitem{KY00}
J.~Kieffer and E.~Yang.
\newblock Grammar based codes: {A} new class of universal lossless source
  codes.
\newblock {\em IEEE Transactions on Information Theory}, 46(3):737---754, May
  2000.

\bibitem{HY03}
D.~He and E~Yang.
\newblock On the universality of grammar-based codes for sources with countably
  infinite alphabets.
\newblock In {\em Proceedings of IEEE Symposium on Information Theory}, 2003.

\bibitem{JR}
Jeffrey Rosenthal.
\newblock {\em A first look at rigorous probability theory}.
\newblock World Scientific, 2nd edition, 2008.

\bibitem{Kie78}
J.C. Kieffer.
\newblock A unified approach to weak universal source coding.
\newblock {\em IEEE Transactions on Information Theory}, 24(6):674---682,
  November 1978.

\bibitem{HS14}
M.~Hosseini and N.~Santhanam.
\newblock Characterizing the asymptotic per-symbol redundancy of memoryless
  sources over countable alphabets in terms of single-letter marginals, 2014.
\newblock Full version available from arXiv doc id:1404:0062.

\bibitem{osvz04:lim}
A.~Orlitsky, N.P. Santhanam, K.~Viswanathan, and J.~Zhang.
\newblock Limit results on pattern entropy.
\newblock {\em IEEE Transactions on Information Theory}, July 2006.

\end{thebibliography}

\appendix
\newpage
\section{Appendix I}

\newcommand{\asi}[1]{{a^{(#1)}_i}}
\newcommand{\aji}{\asi{j}}
\newcommand{\as}[1]{{a^{(#1)}}}
\newcommand{\aj}{\as{j}}

\newcommand{\amaxi}{{\hat{a}_i}}
\bLemma
\label{seqLemma}
Let $\sets{\aji}$, $1\le j\le k$ be $k$ different sequences with limits $\aj$ respectively. For all $i$, let
\[
\amaxi=\max \aji.
\]
Then the sequence $\sets{\amaxi}$ has a limit and the limit equals $\max \aj$.
\Proof Wolog, let the sequences be such that the limits are
$\as1\ge \as2\ge \ldots \ge \as{k}$. Consider any
$0<\epsilon < \frac{\as1-\as2}2$. Then for all $1\le j \le k$, there
exist $N_j$ such that for all $n\ge N_j$, $|\aji - \aj| \le \epsilon$. Let 
$N=\max N_j$. We now have that for all $i\ge N$, 
\[
\amaxi = \max \aji = \asi1,
\]
and therefore, the sequence $\sets{\amaxi}$ has a limit, and is equal
to $\as1= \max_{1\le j\le k} \lim_{i\to\infty} \aji$.
\eLemma

\section{Appendix II}
He and Yang~\cite{HY03} considered compressing stationary
ergodic sources with grammar based codes. Let $\cL= \{L_1, L_2...\}$ be bijections from $\naturals\to\naturals$,
and let $q$ be the probabilities corresponding to Elias
encoding of integers, namely $q(i)=1/2^{1+\floor{\log i}+2\floor{\log(1+\floor{\log i})}}$
over $\naturals$. For any stationary ergodic source $p$
such that there exists $L^*\in\cL$ satisfying
\[
E_p \log \frac1{q(L^*(X_1))}\le \infty,
\]
Theorem 2 in~\cite{HY03} shows that for all processes with finite entropy rate, there is a measure
$\phi$ over infinite sequences of $\naturals$ achieving
\begin{equation}
  \label{eq:hy}
\limsup_{n\to\infty} \bigg( \frac1n\log \frac1{\phi(X_1\upto X_n)}- H_p(X_{b+1}| X_1^b) \bigg)
\le 5\sum_{x: L(x)\geq b} p(x)\log \frac1{q(L^*(x))},
\end{equation}
from which one can obtain by letting $b\to \infty$ that for all $p$
\[
\limsup_{n\to\infty}  \log \frac1{\phi(X_1\upto X_n)}
\le H_p,
\]
where $H_p$ is the entropy rate of the underlying source.

It is tempting to try and adapt the proof for our case, but the result
above is not strong enough for this. To see this, first note that
in the iid case, $H_p(X_{b+1}|X_1^b)=H_p$, the entropy of the marginal, 
so we can simplify~\eqref{eq:hy} to yield for all $p\in\cP$, 
\[
\limsup_{n\to\infty} \bigg( \frac1n\log \frac1{\phi(X_1\upto X_n)}- H_p \bigg)
\le 5\sum_{x: L(x)\geq b} p(x)\log \frac1{q(L^*(x))}
\]
To make the right side look ``like'' the tail redundancy, we could take the
supremum of both sides above and then a limit as $b\to\infty$ to yield
\[
\sup_{p\in\cP}\limsup_{n\to\infty} \bigg( \frac1n\log \frac1{\phi(X_1\upto X_n)}- H_p \bigg)
\le \lim_{b\to\infty}\sup_{p\in\cP}5\sum_{x: L(x)\geq b} p(x)\log \frac1{q(L^*(x))}.
\]
This is quite different from the redundancy claim for the \iid case
since our claim for the class $\cP$ of \iid sources is that for 
\[
\limsup_{n\to\infty} \sup_{p\in\cP} \frac1n E\log \frac{p(X^n)}{q(X^n)} \le \cT(\cP),
\]
since in general, for any function $f(p,n)$, we only have 
\[
\sup_{p} \limsup_{n\to\infty} f(p,n) \le \limsup_{n\to\infty} \sup_{p} f(p,n).
\]
and usually, we cannot further refine the above inequality to an
equality. To see this, let $\cP$ be a class of all finite support
distributions, and let $f(p)= |\text{support}(p)|/n.$ The left side is
$0$, while the right is infinite.

In fact, we could try adapt the proof of~\eqref{eq:hy} before taking
the $\limsup$ over $n\to\infty$. However, one of the terms in the
upper bound for finite $n$ is
\[
  3|\# \text{of distinct symbols in } X_1\upto X_n|,
\]
which has to grow sublinearly with $n$ for~\cite[Theorem 2]{HY03} to
hold. For a fixed source $p$, this is indeed true. However, even for
\iid classes $\cP$,
\[
  \sup_p E|\#\text{ distinct symbols in} X_1\upto X_n|
\]
cannot be upper bounded with a non-trivial bound ($< n$) in
general. To see why, from~\cite{osvz04:lim} we have that the
\[
E|\#\text{ distinct symbols in } X_1\upto X_n|  \le nH_p /\log n,
\]
where $H_p$ is the entropy of the distribution p, and the above bound
is arbitrarily tight for some sources. Therefore, we can hope to have
a finite upper bound only if $\sup_p H_p < \infty$ as well. However,
this restriction is not appropriate when our goal is to characterize
general classes.

Even in cases where we do have $\sup_p H_p < \infty$, the constants
are still weaker than what is proven in this paper.

\section{Redundancy definitions}
\label{app:rdetails}

Our definitions of redundancy in~\eqref{eq:rn} and~\eqref{eq:r}
conform to the standard definitions, but are more transparent and make
explicit certain nuances that are well known, but implicitly as folk
theorems. As explained in Section~\ref{SecDB}, our definition reveals
the fact that there is nothing to be gained by allowing potentially
inconsistent distributions for different blocklengths while defining
redundancies.  This Appendix proves the equivalence for completeness.

In standard parlance, we usually adopt
\[
  \rho_n \ed \inf_{q_n\in\pnn}\sup_{p\in\cP^n}\frac1n {\mathbb E}\log \frac{p(X^n)}{q_n(X^n)}
\]
where $q_n$ is any distribution over $\naturals^n$ as the length-$n$ redundancy,
while the asymptotic per-symbol redundancy is
\begin{equation}
  \label{std:r}
  \rho \ed \limsup_{n\to\infty} \inf_{q_n\in\pnn}\sup_{p\in\cP^n} \frac1n{\mathbb E}\log \frac{p(X^n)}{q_n(X^n)}
\end{equation}
In this appendix, we show that
\[
  \rho_n = \inf_{q\in\pni}\sup_{p\in\cP^n}\frac1n {\mathbb E}\log \frac{p(X^n)}{q(X^n)} =R_n
\]
and that
\[
  \rho = \inf_{q\in\pni} \limsup_{n\to\infty}\sup_{p\in\cP^n} {\mathbb E}\log \frac{p(X^n)}{q(X^n)} = R,
\]
where $R_n$ and $R$ are our definitions from~\eqref{eq:rn} and~\eqref{eq:r}.

\bClaim
\label{claim:rn}
For all $n\ge1$, $\rho_n = R_n$.
\Proof
To prove the claim, we simply note that the distribution $q_n$
can be extended to a measure $q_n^*$ by first defining for all
$\x\in\naturals^*$ with $|\x|\le n$
\[
  q_n^*(\x) = \sum_{\substack{\z \in\naturals^n\\\x \preceq \z}} q_n(\x)
\]
and for $\z\in\naturals^*$ with $|\z| = m> n$ using any assignment that enforces
consistency, \ie for each $\x\in\naturals^n$
\[
 \sum_{\substack{\z \in\naturals^m\\\x \preceq \z}} q_n^*(\z) = q_n(\x) 
\]
For example, $q_n^*(\z) = q_n(\x)$ for
$\z = \x \underbrace{1\cdots 1}_{m-n \text{1s}}$, and 0 for all other
$\z$ of length $m$. With this observation, and from the fact that any
probability measure $q$ can be marginalized to yield a distribution
over $\naturals^n$, the claim follows.
\eClaim

\bClaim $\rho = R$.
\Proof From the definition~\eqref{std:r}, we get
for free that
\[
  \rho \le \inf_q \limsup_{n\to\infty}\sup_{p\in\cP^n} {\mathbb E}\log \frac{p(X^n)}{q(X^n)} = R
\]
We will now show that
\[
  R =\inf_q \limsup_{n\to\infty}\sup_{p\in\cP^n} {\mathbb E}\log \frac{p(X^n)}{q(X^n)} \le \rho,
\]
establishing the claim.
From the standard definition~\eqref{std:r}, we know that for each
$\epsilon>0$, there is a sequence $\sets{ q_n\in\pnn : n\ge 1}$ of
distributions over $\naturals^n$ respectively such that
\[
  \limsup_{n\to\infty} \frac1n {\mathbb E}\log \frac{p(X^n)}{q_n(X^n)} < \rho +\epsilon.
\]
Now each $q_n\in\pnn$ can be extended to a probability measure $q_n^*\in\pni$ in
Claim~\ref{claim:rn}. Define the measure $q_\epsilon\in\pni$ by assigning to
each finite sequence $\x$ of natural numbers,
\[
  q_\epsilon(\x) = \sum_{m\ge 1} \frac{q_m^*(\x)}{m(m+1)},
\]
and extending it to a probability measure in $\pni$ on the Borel sigma-algebra 
on the natural product topology in $\naturals^\infty$ as usual.
Now we have
\begin{align*}
  \limsup_{n\to\infty}\frac1n {\mathbb E}\log \frac{p(X^n)}{q_\epsilon(X^n)}
  &\le
    \limsup_{n\to\infty}\frac1n {\mathbb E}
    \log \frac{p(X^n)}{\frac{q_n(X^n)}{n(n+1)}}\\
  &\le
    \limsup_{n\to\infty}\frac1n
    \Paren{{\mathbb E}\log \frac{p(X^n)}{q_n(X^n)} + \log(n(n+1))}
    <\rho +\epsilon,
\end{align*}
thus proving that for all $\epsilon>0$,
\[
  R = \inf_{q\in\pni}\limsup_{n\to\infty}\frac1n {\mathbb E}\log \frac{p(X^n)}{q(X^n)}
< \rho+\epsilon.
\]
The claim follows.
\eClaim

\end{document}